\documentclass[fleqn,letterpaper,11pt]{scrartcl}

\usepackage{times}

\usepackage{cite}

\usepackage[section] {placeins}
\usepackage{graphicx}
\usepackage{amsmath}
\usepackage{fancyhdr}
\usepackage{enumerate}
\usepackage{indentfirst}
\usepackage{xspace}
\usepackage{array}
\usepackage{rotating}
\usepackage{bbold}
\usepackage[
  pdfusetitle,
  pdfkeywords={minerva},
  bookmarks,bookmarksopen,
  bookmarksnumbered,
  pdfpagelabels,
  colorlinks,linkcolor=black,citecolor=black,
  pdfview={Fit},
]
{hyperref}

\newif\ifpdf
\ifx\pdfoutput\undefined
   \pdffalse
\else
   \pdfoutput=1
   \pdftrue
\fi
\ifpdf
   \usepackage{graphicx}
   \usepackage{epstopdf}
\else
  \usepackage{graphicx}
\fi

\newcommand{\minerva}{MINER$\nu$\kern-0.12emA\xspace}

\begin{document}  

.
\begin{flushright}
{\large MINERvA Technical Note No. 080} \\
{\large MINERvA DocDB document 12688} \\
{\large Fermilab Technical Publication FERMILAB-FN-1030-ND}\\
\end{flushright}

\vspace{2ex}
\begin{center} 
{\LARGE Model uncertainties for Valencia RPA effect for MINERvA } \\
\vspace{2ex}
{\large Rik Gran, University of Minnesota Duluth} \\
\vspace{2ex}
\today
\end{center}

\begin{abstract}

This technical note describes the application of the Valencia RPA
multi-nucleon effect and its uncertainty to QE reactions from the
{\small GENIE} neutrino event generator.
The analysis of MINERvA neutrino data in Rodrigues et al. PRL 116 071802
(2016) paper
makes clear the need for an RPA suppression, especially at
very low momentum and energy transfer.  That published analysis does not
constrain the magnitude of the effect; it only tests models with and
without the effect against the data.   Other MINERvA analyses need
an expression of the model uncertainty in the RPA effect.   A
well-described uncertainty can be used for systematics for unfolding,
for model errors in the analysis of non-QE samples, and 
as input for fitting exercises for model testing or constraining
backgrounds.  This prescription takes uncertainties on the
parameters in the Valencia RPA model and adds a (not-as-tight)
constraint from muon capture data.
For MINERvA we apply it as a 2D ($q_0$,$q_3$) weight to {\small GENIE} events,
in lieu of generating a full beyond-Fermi-gas quasielastic events.
Because it is a weight, it can be applied to the generated and fully
Geant4 simulated events used in analysis without a special {\small GENIE} sample.
For some limited uses, it could be cast as a 1D $Q^2$ weight
without much trouble.
This procedure is a suitable starting point for NOvA and DUNE
where the energy dependence is modest,
but probably not adequate for T2K or MicroBooNE.

\end{abstract}

\pagebreak

\tableofcontents

\pagebreak

\section{Introduction}


This paper describes the  method for producing 
the leading RPA suppression effect and
assigning an uncertainty.  It is for use with a neutrino event generator
when charged-current quasielastic (CCQE) events 
are generated with only a simple Fermi Gas model, 
such as {\small GENIE} \cite{Andreopoulos:2009rq}
(version 2.10.2 used in for the figures in this note).
The analysis of the hadronic energy in MINERvA neutrino data in
\cite{Rodrigues:2015hik}
demonstrates the need for an RPA effect on top of a Fermi-gas model of
the nucleus, especially to describe the
suppression at very low momentum transfer.   
There have long been hints in neutrino data of the need for a
suppression effect beyond Pauli blocking, 
since K2K~\cite{Gran:2006jn} and the dawn of
the NuInt workshop era, and even in deuterium bubble chamber data as discussed
in the \cite{Meyer:2016oeg} reanalysis.  
Other MINERvA analyses need
an expression of the model uncertainty in the RPA effect in order to
extract cross sections and interpret neutrino nucleus reaction
models.  However taking ``RPA
on'' vs. ``RPA off'' overestimates the uncertainty -- neither nuclear
theory nor reality would permit zero suppression at very-low $Q^2$.

A well-described uncertainty can be used for systematics for unfolding,
for model errors in the analysis of non-QE samples, and
as input for fitting exercises for model testing or constraining
backgrounds.  It can be used in conjunction with an improved axial form
factor uncertainty such as proposed in \cite{Meyer:2016oeg}, and an
uncertainty estimate for the 2p2h process.  This prescription is one
 component needed to replace the huge, 
default $M_A = 0.99^{+0.25}_{-0.15}$ GeV uncertainty in {\small GENIE} with something
more physical, more detailed, and better targeted.

The Valencia quasielastic with RPA model for neutrino data is described in
\cite{Nieves:2004wx}, with a formal estimate of the
uncertainties in \cite{Valverde:2006zn}.  The model originated with a similar analysis
by some of the same authors and compared to electron scattering data
\cite{Gil:1997jg}.  The resulting model has been compared to muon
capture on nuclei \cite{Nieves:2004wx, Nieves:2017lij}, 
MiniBooNE data \cite{Nieves:2011yp}, and MINERvA data
\cite{Rodrigues:2015hik}, 
with encouraging results.  Further elaboration of the RPA effect 
in the Valencia model was done in \cite{Gran:2013kda}, and the same in
combination with a spectral function in \cite{Nieves:2017lij}.
RPA-type effects in neutrino reactions have long been considered by many authors
\cite{Martini:2016eec,Martini:2009uj,Martini:2010ex,Martini:2011wp,Singh:1992dc,Singh:1993rg,Singh:1998md,Volpe:2000zn,Kolbe:2003ys,Jachowicz:2002rr,
  Pandey:2013cca, Pandey:2014tza, Pandey:2016jju};
interesting comparisons to others and to MINERvA data
are beyond the scope of this work.

The work in this note focuses exclusively on the RPA effect.
Quantitatively we have three samples from the model authors' 
{\small FORTRAN} code:  a no-RPA sample, a relativistic RPA sample, and a
non-relativistic RPA sample.
The ratio of the model with the effect turned
on to a run of the model without any RPA effect at all is effectively
a reweight that we will use in our analyses. 
The following figures illustrate the distortion through the ratio of
the two model variations for the neutrino case, as a function of four momentum transfer
$Q^2$ = $q_3^2 - q_0^2$.  In this paper, we will extensively use the shorthand
for the energy transfer $q_0$ (the energy component of a four-vector, 
also called ``nu'' or $\nu$ by DIS folks or $\omega$ by nucleon/nucleus enthusiasts) 
and the magnitude of the three-momentum transfer $q_3$ (or $|\vec{q}|$ or q).
The anti-neutrino case is basically the same, though a few
complications are more significant and are
described after the neutrino case.


\begin{figure}[htbp]
\begin{center}
\includegraphics[width=7.0cm]{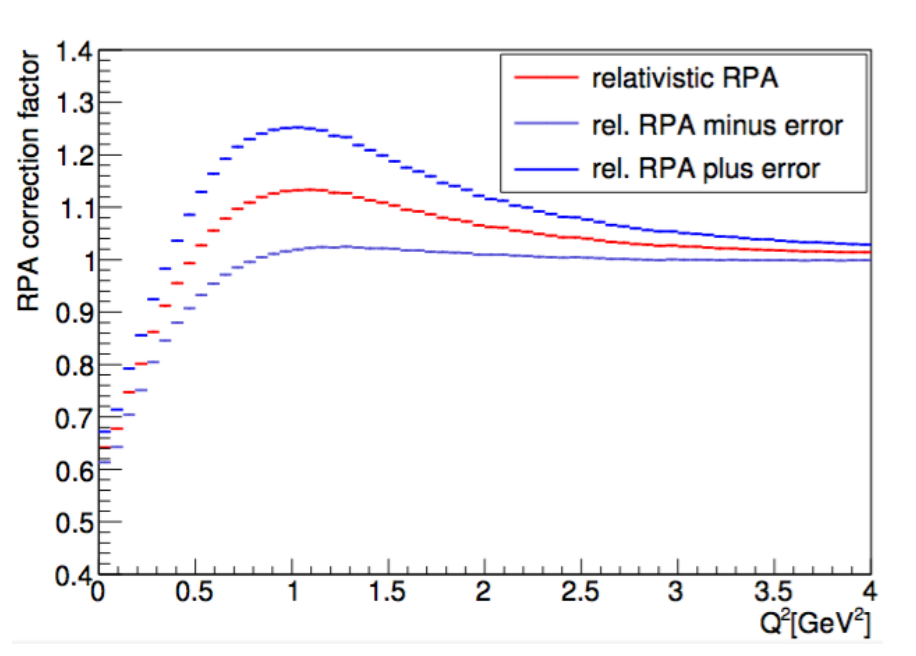}
\includegraphics[width=7.0cm]{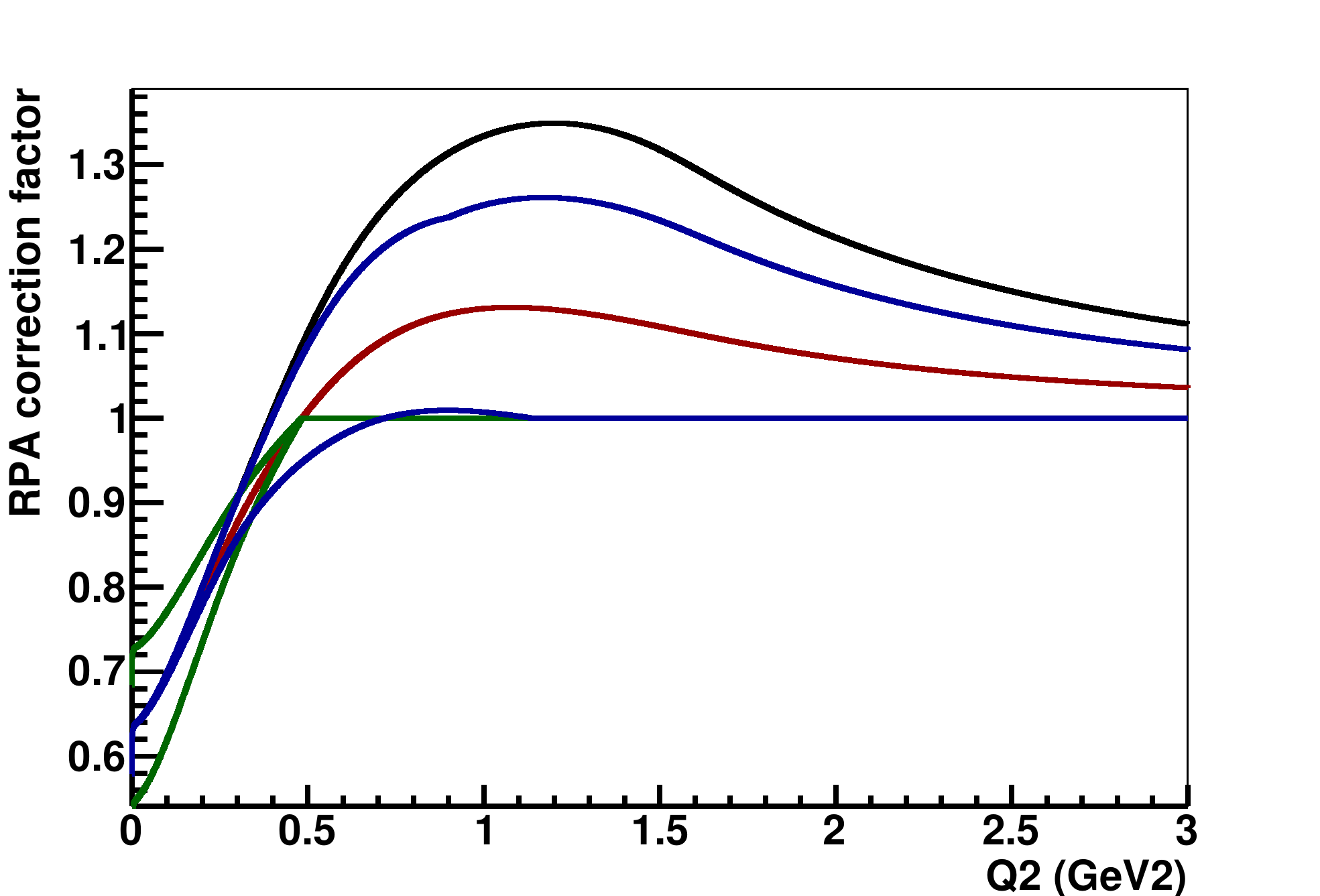}

\parbox{6in}{\caption{Left plot with error band courtesy of Federico Sanchez
    analysis for T2K following the example of \cite{Valverde:2006zn}.
   The central value correction is shown as the red line in both plots
   (from the same model code).
    Right plot, proposed uncertainty for MINERvA (between blue lines)
    tuned using the Sanchez-T2K band at higher momentum
    transfer, and an additional
    uncertainty at low momentum transfer from the muon capture
    constraint (between green lines) described in this note.
The black (top) line on the right is the non-relativistic variant of the RPA
calculation, which is used with the central value to construct the
blue error band lines as described in the text.
\label{fig:uncertainty}}}
\end{center}
\end{figure}

As of this writing, there has been additional effort in refining the
elements of the model and understanding the interplay with spectral
function features \cite{Nieves:2017lij}.  The same RPA code in that
work is used here (but not the spectral function code), and these new comparisons do not
change the interpretation presented here.  The low
$Q^2$ suppression uncertianty based on muon capture data does not need
to be adjusted.

\section{Parameters in the model}

Federico Sanchez and a student at IFAE Barcelona unpacked the
parameters in the Valencia RPA model in order to produce the error band
in the left plot of Fig.~\ref{fig:uncertainty}.  The values of those parameters
were taken as in \cite{Valverde:2006zn} to have 10\% uncertainty.  
Then the calculation was rerun to
produce variations around the central value, which were taken in quadrature.

\begin{table}[htbp]
\begin{center}
\begin{tabular}{ccccc}
parameter & value & & model effect & rough size \\ \hline
$f_{0}^{'(int)}$ & 0.33 & $\pm$0.03 & in $\tau\tau$ potential term & tiny effect \\
$f_{0}^{'(ext)}$ & 0.45 & $\pm$0.05 & in $\tau\tau$ potential term &   tiny effect \\
$C_\rho$ & 2.0 & $\pm$ 0.2 & in $\tau\tau$ potential term & large effect \\
$f$ & 1.00 & $\pm$ 0.10 &  & small effect \\
$f^{\*}$ & 2.13 & $\pm$ 0.21 &  in $\sigma\sigma\tau\tau$ for $\Delta$ & small effect \\
$\Lambda_\pi$ & 1200 & $\pm$ 120 MeV & in $\sigma\sigma\tau\tau$  $V_{long}$ & tiny effect \\

$\Lambda_\rho$ & 2500 & $\pm$ 250 MeV & in $\sigma\sigma\tau\tau$
                                        $V_{trans}$& large effect \\
$g'$ & 0.63 & $\pm$ 0.06 & in $\sigma\sigma\tau\tau$ & large effect
\end{tabular}
\parbox{6in}{\caption{From Federico Sanchez, see slides in MINERvA
    docdb:12621, and \cite{Valverde:2006zn}.
\label{fig:diffxs}}}
\end{center}
\end{table}

The $f$ parameters produce a crossover point around $Q^2$  = 0.5
GeV$^2$ :
when a changed parameter raises the weight (ratio withRPA/without)
above this spot and the effect goes more extremely low below this spot.  The
other systematic parameter shifts raise or lower the weight across all $Q^2$.  The $\Lambda$
parameters do the latter also, but produce no weight in the limit $Q^2 \rightarrow 0$ GeV$^2$.

For the enhancement at high $Q^2$, above the crossover point, all
these parameter uncertainties produce the same kind of distortion.
Our implementation, therefore, does not need each individual
parameter, but only one parameter whose 1$\sigma$ is the sum of the
effects in quadrature, which is what was shown in Fig.~\ref{fig:uncertainty}
left plot.  We will construct the equivalent magnitude from simpler ingredients.

Below the crossover point we would need a more sophisticated
treatment to capture the uncertainty.  
However there is an additional data-driven uncertainty that
is not yet accounted for in Sanchez' analysis, yet appears to be the
largest effect.  This is described in the next section, and can
supply all the low $Q^2$ uncertainty we need.

\section{Additional uncertainty at very low momentum transfer}

The success of the RPA multi-nucleon effect has been tested \cite{Nieves:2004wx} against
data for muon capture on nuclei.   The Feynman diagram for this
process is equivalent to the muon neutrino CCQE process in nuclei we are
studying in MINERvA, but the data for the process are necessarily
limited to very-low $Q^2$.  
The Valencia authors make a comparison of the rate for muon capture to
their model with and without RPA.  We will use that
data-driven observation to create a
second, independent uncertainty band.

\begin{figure}[htbp!]
\begin{center}
\includegraphics[width=15.0cm]{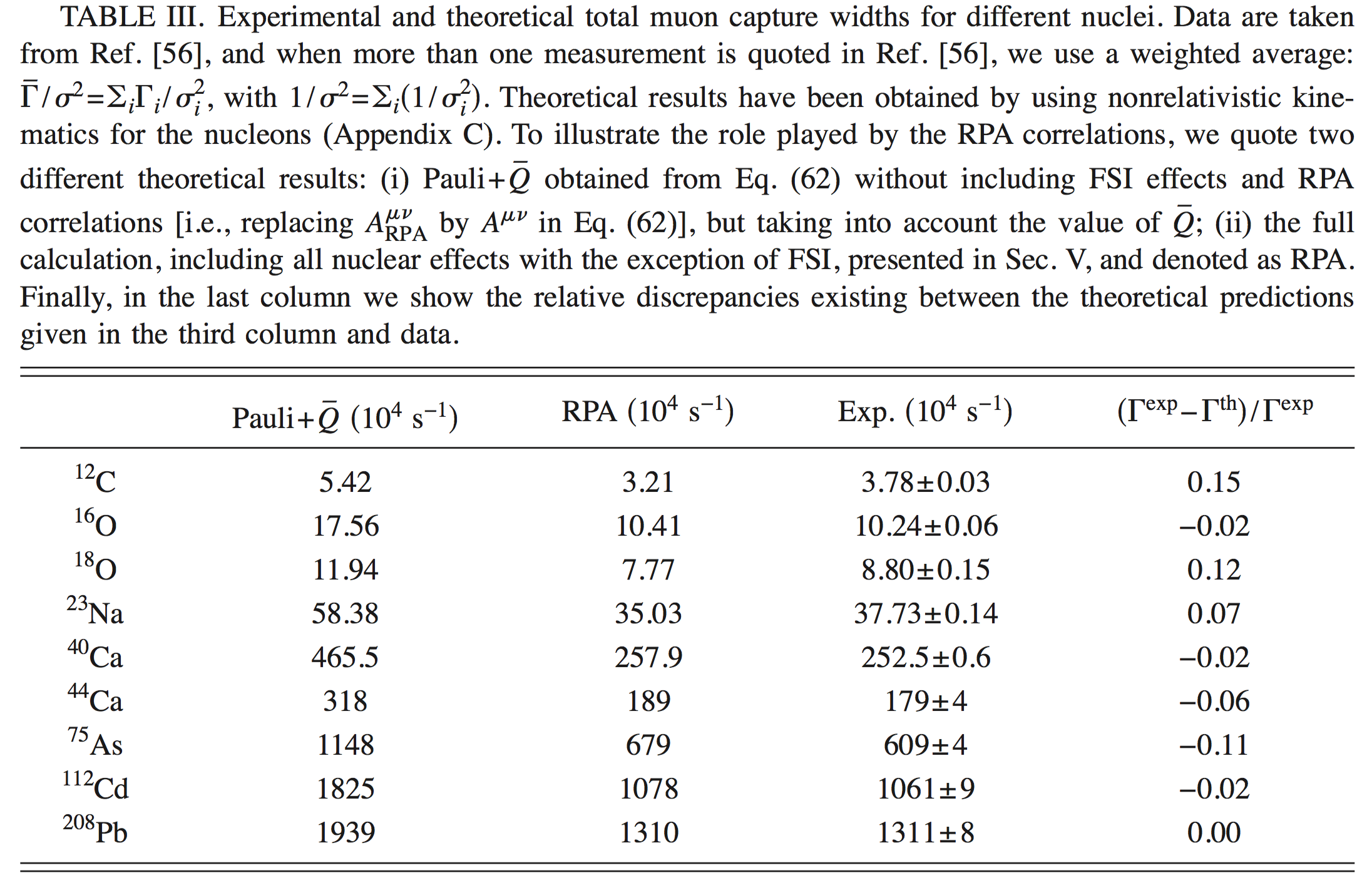}
\includegraphics[width=15.0cm]{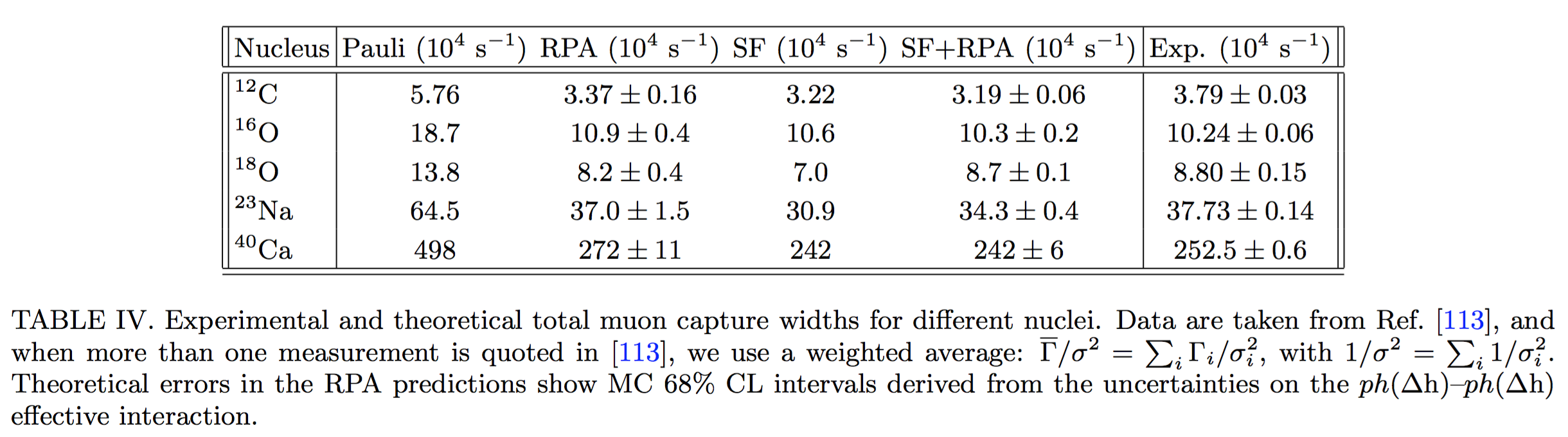}

\parbox{6in}{\label{tab:nievestable}\caption{Screenshot of Table III
    and caption from Nieves {\em et al.} paper
    \cite{Nieves:2004wx} (upper table) and the improved version from
    Table IV of Nieves
    and Sobczyk \cite{Nieves:2017lij} (lower table) for a subset of the same nuclei 
    showing the improved agreement with muon
    capture data when the RPA and Spectral Function effects are included.   The remaining
    disagreement drives our uncertainty estimate.
}}
\end{center}
\end{figure}

This figure \ref{tab:nievestable} (really a screenshot of Table III from their
\cite{Nieves:2004wx} paper and Table IV of \cite{Nieves:2017lij}) 
shows the model with RPA clearly describes
the muon capture rate, and the one without does not describe it well
at all.

The fractional discrepancy is given
in the right column, but it makes more
convenient reweighting code for us to recast that as an 
{\em uncertainty in the suppression from nominal}. 
For example, the suppression factor
for carbon in the table is 3.21/5.42 = 0.6 = (1.0 $-$ 0.4) in for RPA model to
default model, 
but 3.78/5.42 = 0.7 = (1.0 $-$ 0.3) for
data to default model.  This 0.3 compared to 0.4 is 25\% less suppression.  We can
implement this as scaling (pseduo-math): 
default $\pm$ 0.25 ( 1.0 $-$ default ) ;  
\\so for carbon 0.6 $\pm$ 0.25 $\times$ (1.0 $-$ 0.6) = 0.7 or 0.5,
\\producing the green band on the right plot in Fig.~\ref{fig:uncertainty}.

Overall, the model gives a suppression factor of 0.6 for practically
all nuclei, as does the average of the data for these nuclei.  
[A spreadsheet is included in
the MINERvA docdb for this technical note.]  
The standard deviation of the remaining discrepancy between model and
data
for the nuclei in the table 
is 10\%, and could be used as an uncertainty.  However, the carbon
nucleus is essential to our physics program, and has the largest
discrepancy, so we should take that larger value to represent our
1$\sigma$ uncertainty for MINERvA.   
There are subtle differences between the RPA effect for different
nuclei, described in Sec.~\ref{sec:othernuclei}, we will apply the
uncertainty prescription to a base central value specific to carbon,
oxygen, iron, and lead, rather than reusing carbon for all of them.


There is some difference in the RPA effect for anti-neutrinos, 
which are exaggerated by an energy dependence, which is described
in Sec.~\ref{sec:energydependence}.  It is driven by the vector-axial
interference term, but is significant only at momentum
transfers near the upper kinematic limits.  It may cause artifacts when the high momentum transfer
weights from the Valencia model are applied to events generated from
another model like {\small GENIE}.   For MINERvA, we have simplified our treatment
of the weight as described in Sec.~\ref{sec:modelskew}.

Finally, there is negligible difference between the predicted
suppression for the $v_e$ and $v_\mu$ cases, and we use the latter for
all neutrino reactions.  The neutral current case is not currently
considered, but should also show the effect, as it was obviously there also for
electron scattering \cite{Gil:1997jg}.  

The RPA effect at low $Q^2$ can be though of as a screening effect.
The spectral function effect is more like a transfer of cross section
strength to a high energy transfer tail, which can produce a
distortion similar in some ways to the RPA effect.  A
followup study of nuclear effects \cite{Nieves:2017lij} by Juan Nieves
and Joanna Sobczyk incorporate spectral function like elements into
the previous analysis, and remake comparisons to muon capture data in
their Table IV.
The variations of RPA, Spectral Function, and the two combined cluster
closer together than the discrepancy with the data for Carbon.
Therefore the prescription for a 25\% uncertainty presented in this
section still accomodates the tension between these models and the
data, and is suitable for use without modification.  Other aspects of
the spectral function's predicted energy and momentum transfer spectra are distorted more,
and in ways not described by the RPA calculation.

\section{Implementation of an uncertainty band}
\label{sec:implementation}

Repeating what was shown in Fig.~\ref{fig:uncertainty}, 
the Valencia model has both a relativistic calculation, which is our
choice for the central value RPA effect, and also a non-relativistic
calculation, and of course a CCQE calculation without RPA.   
A histogram file, saved from running their model
code directly, includes all three calculations.   
Ratios are made from
these calculations at $E_\nu$ = 20 GeV
for both 1D $Q^2$ distribution and
on a 2D grid for momentum and energy transfers from
0.0 to 5.0 GeV.  The differences between using the 1D or 2D inputs are
described over the next two sections.

To produce a $+1\sigma$ systematic that makes the enhancement bigger
for most $Q^2$, we produce a weight that is 60\% 
of the way from the central value RPA effect to the
non-relativistic one.   This fraction is increased smoothly
from  $Q^2= 0.9$ GeV$^2$ down to lower $Q^2$, increasing the uncertainty to match
the full non-relativistic prediction at a lower $Q^2$ of 0.4 GeV$^2$.  This
one-parameter magnitude is a good description of the band from the
Sanchez/T2K analysis and is easy to implement.   

The $-1\sigma$ is made by going that far
to the low side of the central value, but truncating the weight and
not allowing it to go below 1.0 at high $Q^2$.   The structure of the
model does not allow a suppression at high $Q^2$, so this
implementation also does not permit it.   Note, this
feature means we can not form a meaningful $-2\sigma$ uncertainty,
which requires special treatment in a many-universe method analysis.

Independently, a second set of $\pm1\sigma$ weights are produced in the low momentum
transfer regions (below 1.0 GeV) that reduce or increase the amount of
suppression by 25\%.   These yield a data-driven error from the muon
capture comparison, and are
assumed to be uncorrelated with the uncertainty in the enhancement.


\section{The implementation works also for a 2D weight}
\label{sec:2Dweight}

Sanchez and others in the past have done their 
analyses in one dimension $Q^2$, but the RPA effect in
the model has substructure in energy transfer $q_0$ (also called ``nu'' $\nu$ or
$\omega$) and three-momentum transfer $q_3$ (or $|\vec{q}|$ or q).
In the Valencia calculation, 
the suppression appears to be more a function of $q_0$ than $Q^2$.
In contrast, the enhancement has the effect of slightly shifting the QE peak to
lower $q_0$ in addition to the broad $Q^2$ trend.

We capture these effects by implementing the CCQE reweight in two
dimensions, using a lookup table saved as a TH2D Root histogram.   The
weight is made from the ratio of the CCQE with RPA / CCQE without RPA,
both with the Valencia model, and so only captures the RPA effect.  
Other aspects that may
make the Valencia model different than {\small GENIE}, such as their choice of
M$_A$ or the use of a local Fermi gas are not part of the weight, but
are visible in this analysis.

Two important points.  The prescription in the previous section
{\em is applied at each point in a 2D space} of
($q_0$,$q_3$), not $Q^2$, but will yield the required $Q^2$ distortion
for analyses that integrate the cross section to produce the $Q^2$
distribution.  There are differences in ($q_0$,$q_3$) 
between the {\small GENIE} events to be
weighted and the Valencia cross section from which we obtain the
weights that require a slight variation on the prescription.

The next subsections illustrate the 2D weight from the Valencia model,
and are designed to allow you to flip-book the neutrino and
anti-neutrino weights to see how similar they are.

For technical completeness, these are the same 20.0 GeV neutrino energy
calculations from the Valencia authors' {\small FORTRAN} code.
They are binned finely with 1000 bins per GeV in momentum and energy
transfer up to 5.0 GeV.  With such fine binning, no interpolation is
needed or used.  However, the resulting filesize would typically be of
megabytes when the full 5.0 GeV range is used (for each nu and
anti-nu, for each nucleus).  For running 
under MINERvA production-style Grid Computing conditions, histograms
are made only up to 3.0 GeV.  More coarsely binned histograms, with an
interpolation could also be made, if Grid bandwidth or memory restrictions
require it.
The 2D histograms are
the most natural basis for the calculation.  The $Q^2$ curves from the
Valencia model shown in this paper are made directly integrating the double-differential
cross sections.

\clearpage
\pagebreak

\subsection{Neutrino 2D ratio used as central value weight}

\begin{figure}[htbp]
\begin{center}
\includegraphics[width=7.0cm]{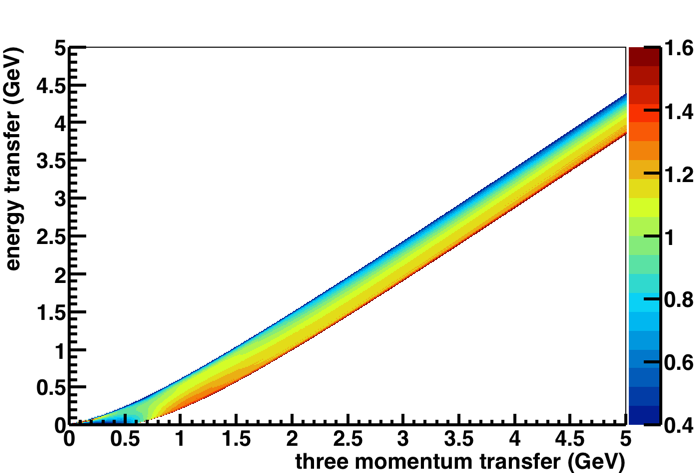}
\includegraphics[width=7.0cm]{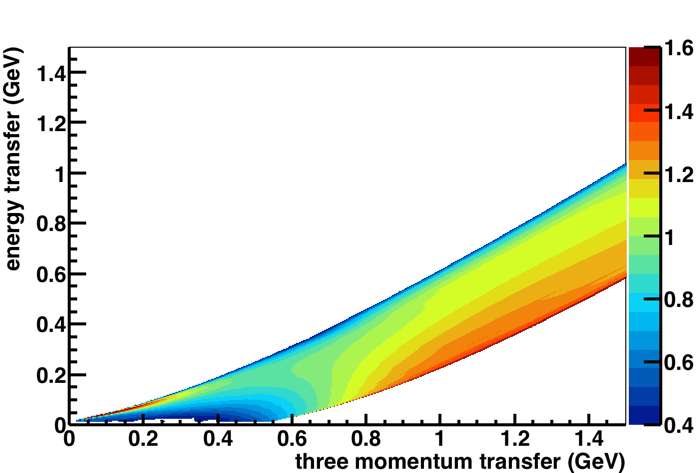}
\includegraphics[width=5.0cm]{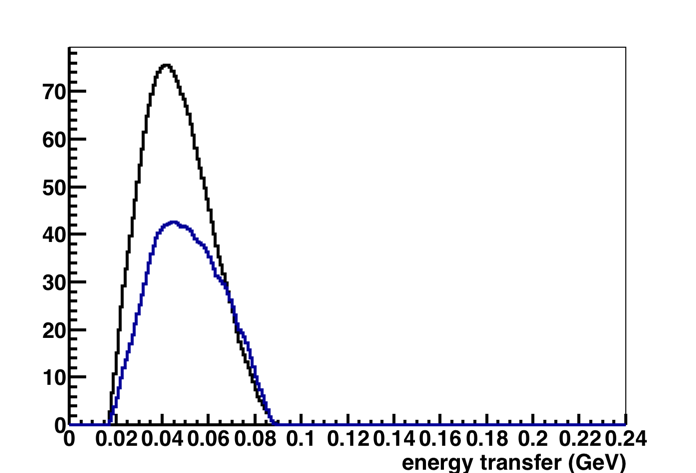}
\includegraphics[width=5.0cm]{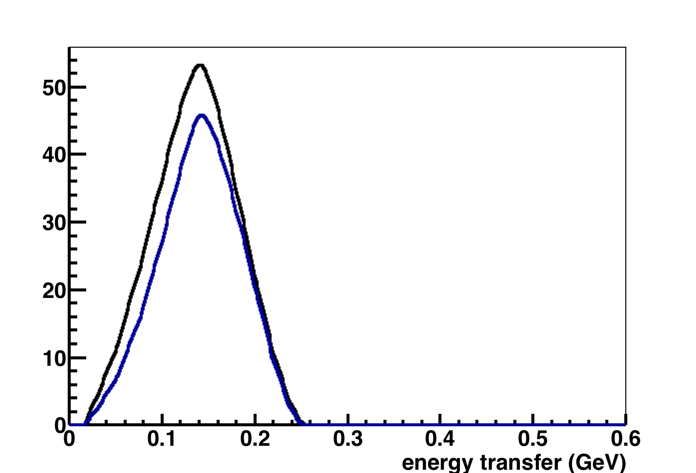}
\includegraphics[width=5.0cm]{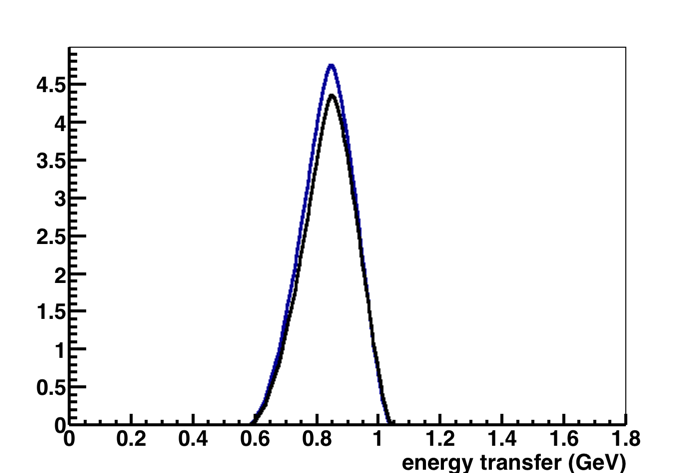}
\parbox{6in}{\caption{Upper figures: the central value RPA weight from the ratio of
    the Valencia QE model with RPA / without RPA. The right plot is the same,
    but zoomed in on the low momentum-transfer region.
   Lower figures:  slices of constant three-momentum transfer at 
  $q_3$ = 0.2, 0.5, and 1.5 GeV, blue is the version with RPA.
    You can imagine the lower figures illustrate that
    approximately the QE distribution is centered on a line that runs
    through the middle of the 2D band.
    The electronic version of this note may allow you to flip pages
    between this figure and the anti-neutrino version.
\label{fig:baseratio}}}
\end{center}
\end{figure}

The low momentum transfer region shows a suppression that reduces
energy transfers on the low side of each peak more than on the high
side of each peak.  By $q_3 = 1.5$ GeV, the enhancement (blue curve is
higher than the black) is approximately
but not perfectly the same shape as the original peak, enhanced and 
relatively wider and shifted slightly to lower energy transfers.

\clearpage
\pagebreak

\subsection{Anti-neutrino 2D ratio used as central value weight}

\begin{figure}[htbp]
\begin{center}
\includegraphics[width=7.0cm]{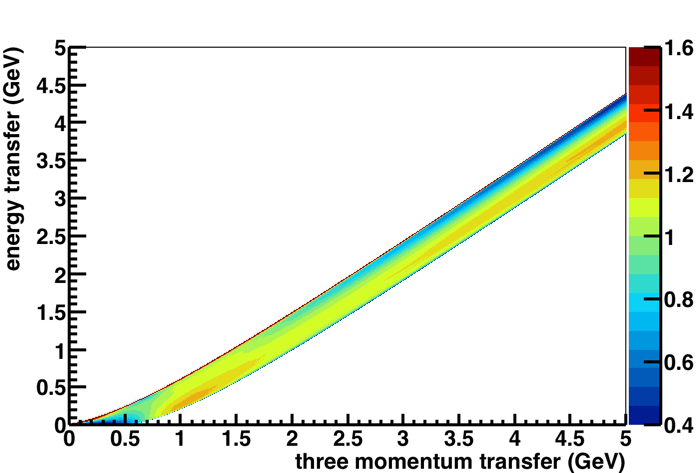}
\includegraphics[width=7.0cm]{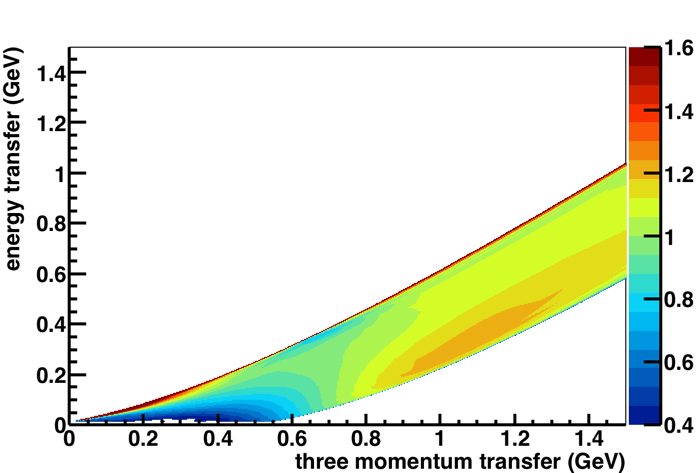}
\includegraphics[width=5.0cm]{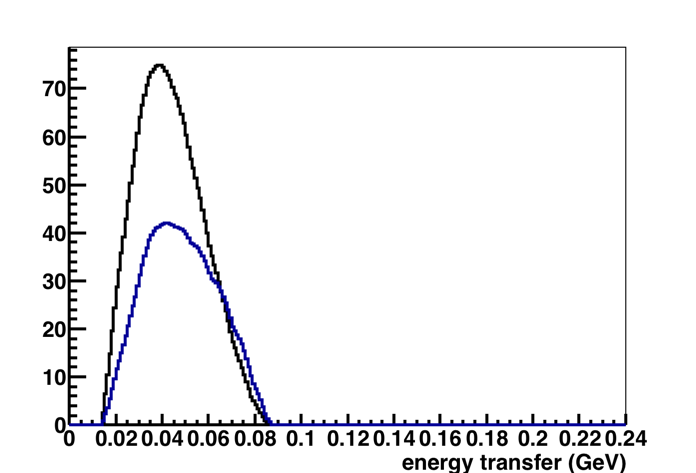}
\includegraphics[width=5.0cm]{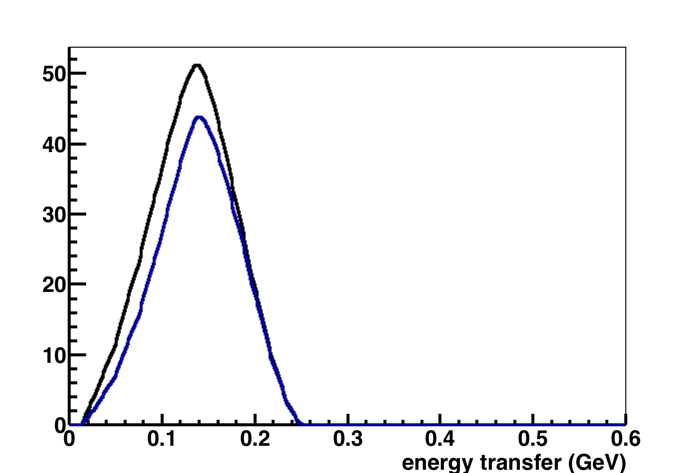}
\includegraphics[width=5.0cm]{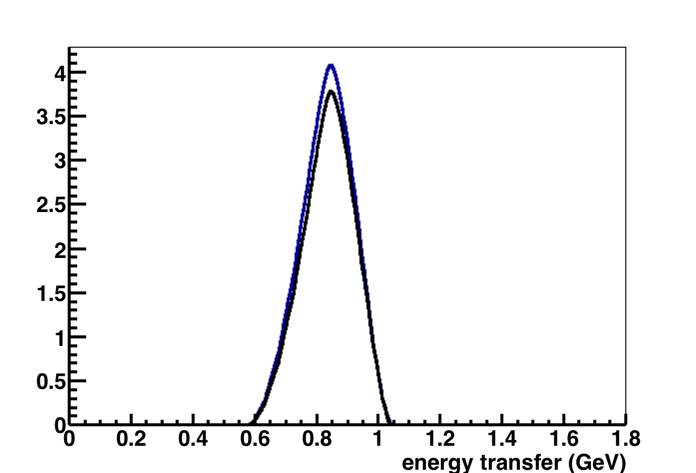}
\parbox{6in}{\caption{Upper figures: the central value RPA weight from the ratio of
    the Valencia QE model with RPA / without RPA. The right plot is the same,
    but zoomed in on the low momentum-transfer region.
   Lower figures:  slices of constant three-momentum transfer at 
  $q_3$ = 0.2, 0.5, and 1.5 GeV, blue is the version with RPA.
    You can imagine the lower figures illustrate that
    approximately the QE distribution is centered on a line that runs
    through the middle of the 2D band.
    The electronic version of this note may allow you to flip pages
    between this figure and the neutrino version.
\label{fig:anuratio}}}
\end{center}
\end{figure}

The anti-neutrino trends have the same magnitude up to about 1.5 GeV.
The low momentum transfer region shows a suppression that reduces
energy transfers on the low side of each peak more than on the high
side of each peak.  By $q_3 = 1.5$ GeV, the enhancement (blue curve is
higher than the black) is approximately
but not perfectly the same shape as the original peak, enhanced and 
relatively wider and shifted slightly to lower energy transfers.
The magnitude of the effect is slightly different, presumably
following the vector-axial interference term's significance as it
changes sign.

\clearpage
\pagebreak

\section{Model Skew Workarounds}
\label{sec:modelskew}

Because we are applying an ``optional'' feature of one model on top of another
model's base predictions, we will be afflicted by (and need to
workaround) cases where the two base predictions disagree.
(Its only optional from the code's perspective.  Its not optional physics.)
Most relevant, we are faced with applying a weight to some {\small GENIE}
events that have no analog in the Valencia model, and vice versa.
A second concern, the ($q_0$ or $W$) peaks of the two distributions are not in
identical locations nor do they have identical shapes.
We can do something sensible with these; its a classic
case of not letting perfect be the enemy of the good.   If MINERvA had the
ability in 2016 to cleanly use a consistent base model, we would, but
that is not a luxury we have until 2017.  Instead, we are bootstrapping to a
2017/2018 model future, learning along the way.

The following two 1D plots for 5 GeV anti-neutrino carbon quasielastic
reactions show the effects better than the 2D plots.  The Valencia
model is the no-RPA, yes local Fermi-gas version; the {\small GENIE} model is
the default global Fermi-gas.  (There is no global Fermi-gas version
of the Valencia model).  The left is at the lowest $q_3$ = 0.1 MeV, the
right is higher $q_3$ = 1.5 GeV.

\begin{figure}[htbp]
\begin{center}
\includegraphics[width=7.0cm]{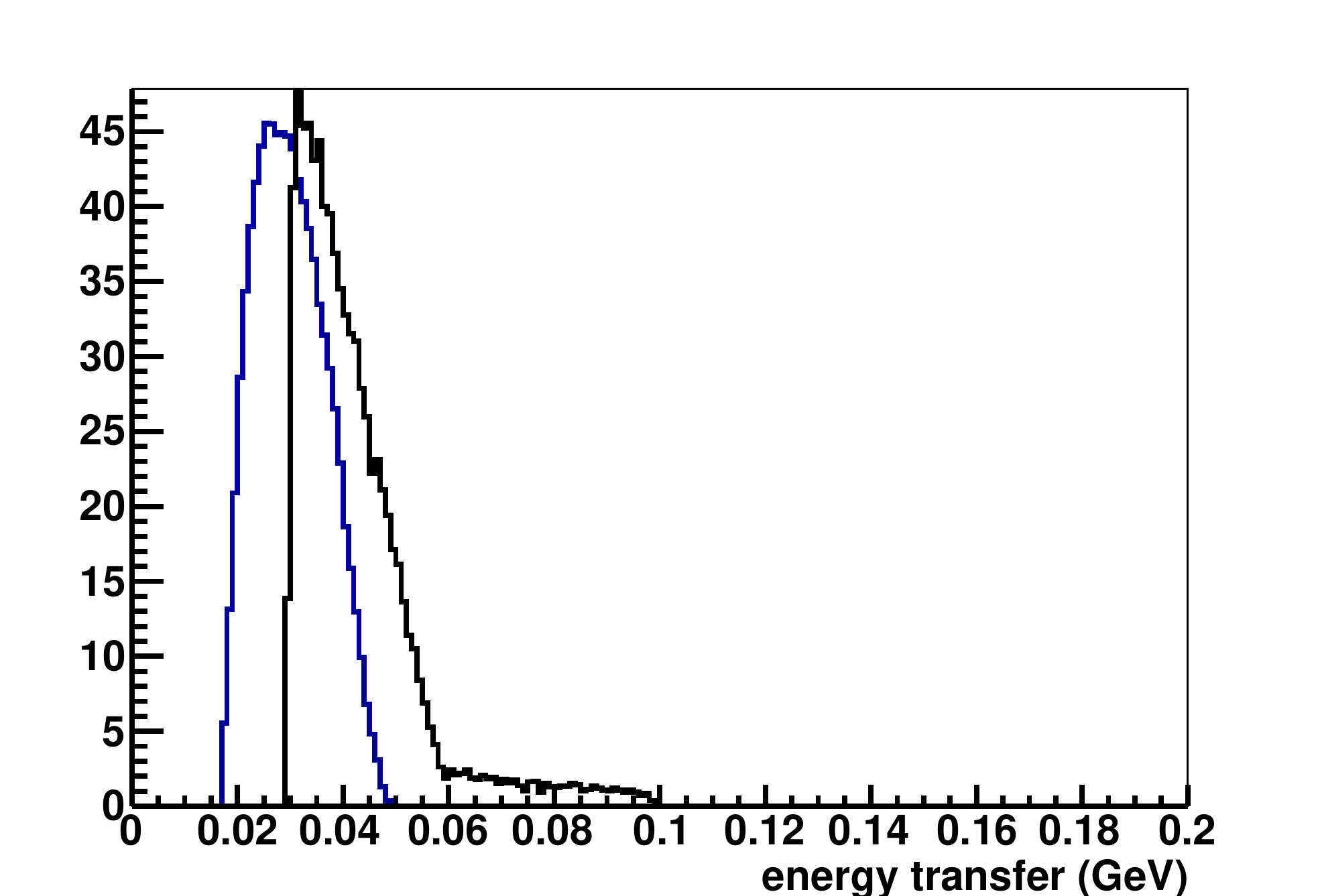}
\includegraphics[width=7.0cm]{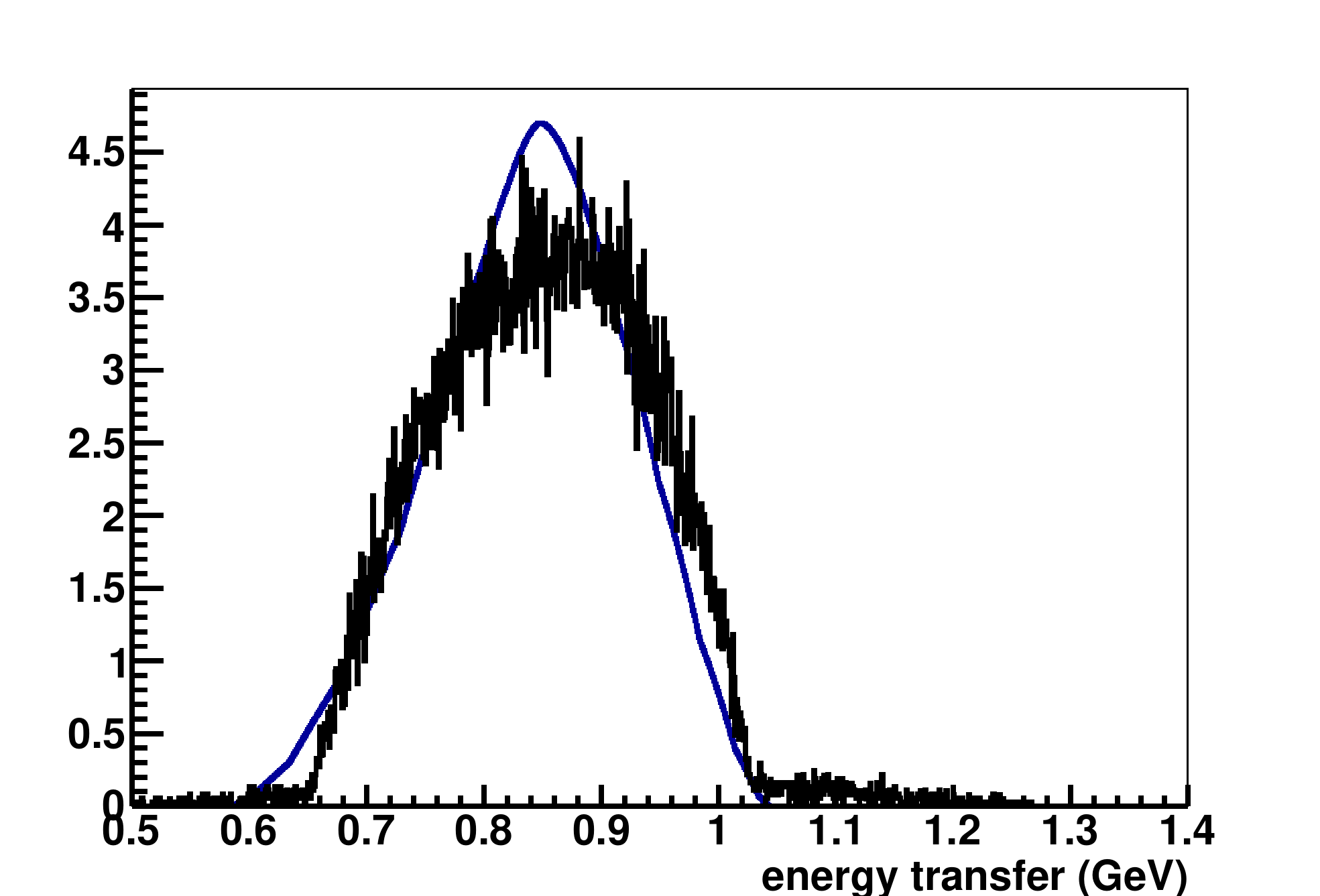}
\parbox{6in}{\caption{Comparison of Valencia (Blue) and {\small GENIE} (Black)
    cross sections in different slices of $q_3$ from $E_\nu = 5$ GeV
    calculations. 
    {\small GENIE} is
    consistently $\sim$10 MeV higher than the Valencia model, an
    offset due to thresholds and energy
    conservation differences.  Also, the {\small GENIE} model has a distinct
    tail to higher energy transfers
   due to the Bodek Ritchie ``spectral-function''-like addition.
   Shape differences are also due to the Valencia use of a local Fermi-gas.
\label{fig:oneDplots}}}
\end{center}
\end{figure}

The pattern is a systematic shift of the cross section up by about 10
MeV; {\small GENIE} is the higher one. 
This is probably as simple as an overall offset
due to different assumptions about removal energy.  In effect, {\small GENIE}
produces a higher threshold of 41 MeV for carbon, a combination of
removal energy (25 MeV) and a Pauli blocking artifact, 
so events below that threshold do not contribute to the cross section
or event rate.   Valencia assumes something more like 17 MeV for
neutrinos and 14 MeV for anti-neutrinos, coming from using the
Q-value for the charged current reaction in the initial energy
conservation step.  Its
possible there are complications due to the local Fermi-gas nature
too, which allow different removal energies for different locations in
the nucleus.

\subsection{enhanced prescription item one: energy transfer offset}

So the prescription described earlier has five more ingredients.
We have engineered that the RPA weight given to a {\small GENIE} event comes from
Valencia kinematics at the same $q_3$ but 0.010 GeV lower energy transfer.

\subsection{enhanced prescription item two: momentum transfer offset}

The second feature is that {\small GENIE} has a wider distribution, due in part
or in most to the Bodek Ritchie tail designed to kinematically mimic the
SRC/spectral function feature of the nucleon momentum distribution.
These events do not have a corresponding weight in the Valencia
model, though also they are a minority of events.

We identify these events, below 0.15 GeV energy and momentum transfer,
as having no weight (zero weight) in the 2D histogram.
Instead of eliminating them (weight of zero) or doing nothing (weight
of 1.0), we assign them the weight from the
Valencia model from the same energy transfer but a momentum transfer
0.150 GeV higher.

\subsection{enhanced prescription item three: switch to parameterized weight}

The underlying shapes and peak locations are different, 
even the middle of the peak at higher momentum transfers.   
The Valencia peak is narrower, an expected outcome of using a
local-Fermi gas.
Applying a weighting from one shape
that tries to enhance and shift the cross section to
another whose peak is different could have a detrimental effect, 
suppressing the cross section instead of enhancing it.
In this situation, it is easiest to
choose between preserving the few MeV shift or the enhancement of the
integrated cross section, whichever is more important for the physics studies.

For MINERvA at high momentum transfers
(i.e. relatively energetic hadron systems) preserving the integrated
cross section is probably more important.
The 2D histogram will be used only  from $0.0 < Q^2 < 3.0$ GeV$^2$, 
roughly equivalently $0.0 < q_3 < 2.3$ GeV.
The switch will also happen for $q_3 > 3.0$ GeV, which is a tiny part
of parameter space not already covered, but allows us to optimally
squeeze the input histogram size.
Any events with higher momentum transfer will be assigned a weight 
from a polynomial parameterization $r(Q^2)$ up to 
$Q^2 = 9$ GeV$^2$ (or $q_3 \sim 5$ GeV).
Finally, for the rare QE events at higher momentum transfers, we will
assign no weight = 
1.00 $^{+0.03}_{-0.00}$.

\subsection{enhanced prescription item four:  enhance the high
  momentum transfer uncertainty}

At high momentum transfers there are artifacts of energy and $Q^2$
dependence in this method that are larger than the difference between
the relativistic and non-relativisic calculations.  These artifacts
could be as big as 2\%.    As the
prescription described here leads to a value of the high $Q^2$
enhancement of around 1.02$^{+0.03}_{-0.02}$ at roughly $Q^2=3$ GeV$^2$,
the resulting uncertainty band will be smaller than these artifacts.
We inflate the positive error to remain 0.03 above the central value
and the negative error to remain pinned to a weight of 1.0.
Actually, the anti-neutrinos have larger artifacts than the neutrinos,
and we could use different errors in this spot.  However, this is a
situation where these model (and model-implementation) uncertainties
will be dwarfed by the uncertainty in the axial form factor, and we
need a concrete procedure more than we need to worry about the error
on an error.

\subsection{enhanced prescription item five:  limit enhancement to 2.0}

There are kinematics along the upper edge of the QE band where the RPA
version of the model puts events where the no-RPA version has very few
events.  This leads to a narrow region that gets excessivey large weights.
If it was applied to just the Valencia model, the net effect would be
right.  Because those weights are applied to a different default
model, it can produce an anomalous absolute event rate at some
kinematics.  To prevent this, weights above 2.0 are forced back to
exactly 2.0.  

\subsection{one more comment}

The fundamental prescription in this technical note can be used
to provide an uncertainty band even if the Valencia model itself is
the base model being considered.  For example, a version of the model
is included in GENIE 2.12 in a form that includes explicit generation
of proton kinematics, which this reweighting procedure can not do.   
With no skew between models, 
not all the enhanced prescription steps are necessary.  Possibly only
item four above is desireable.  As of this writing, GENIE 2.12 does
not generate an uncertainty band like is described here.


\clearpage
\pagebreak

\section{Q2 vs energy transfer weights}
\label{sec:q2vs2D}

Historically, analysis of neutrino data has centered on $Q^2$
distributions, and things like an RPA suppression has been expressed
as a function $r(Q^2)$ .   The Valencia calculation features a
suppression that is arguably preserves more detail in 
$r(q_0,q_3)$ form.  The latter comes at the cost of a more
computationally intense 2D weight.  Is the difference observable, and does it
matter?  Probably.

These two plots show the distortion of {\small GENIE} events using two
different RPA weighting methods.  The left one using the method we propose
for MINERvA; it is the central value weight actually applied after all
details of the implementation described previously.  
The right one uses a polynomial $r(Q^2)$ fit to the same
Valencia model projected onto Q2, then applied to the same {\small GENIE} events.  

\begin{figure}[htbp]
\begin{center}
\includegraphics[width=7.0cm]{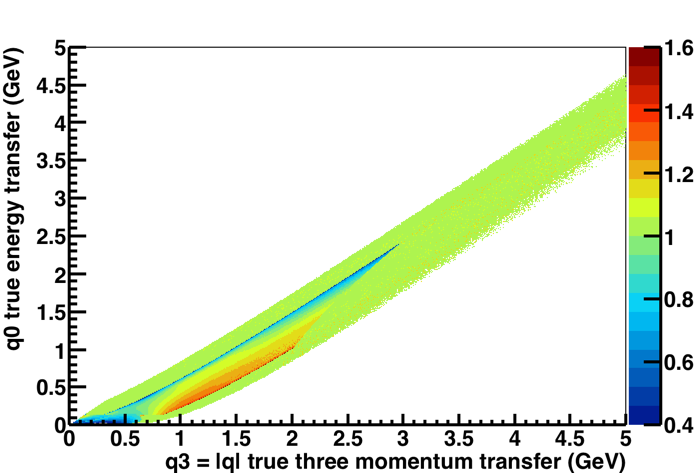}
\includegraphics[width=7.0cm]{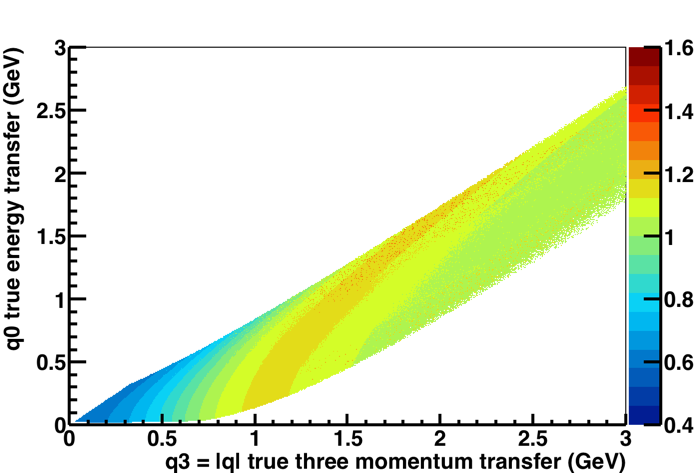}
\parbox{6in}{\caption{The 2D histogram-ratio-based weight
    r($q_0,q_3$) switching to the parameterized r($Q^2$) at $q_3$ = 2.3
    GeV as proposed compared
    to the same parameterized r($Q^2$) weight built from the
    same Valencia model projected onto $Q^2$ but used over the whole
    range.
    The left plot has a radically
    different structure, even though the two weights produce nearly
    the same distortion of the $Q^2$ spectrum.  This could be
    observable in the MINERvA low-recoil \cite{Rodrigues:2015hik} analysis,
    but essentially the same in a $Q^2$ analysis.
\label{fig:qqparameterization}}}
\end{center}
\end{figure}

Notice the suppression has a radically different effect as a function
of $q_0$ for momentum transfers $q_3 < 0.5$ GeV.   This is
particularly
interesting kinematics for the MINERvA ``low-recoil'' analysis
\cite{Rodrigues:2015hik} observables.   
To use the $r(Q^2)$ weight, even if derived from the Valencia
calculation, would not at all be keeping in the spirit of the physics
of that calculation.  In particular, the $r(Q^2)$ 
version systematically suppresses the
QE in the dip region, which is the opposite of what the MINERvA data \cite{Rodrigues:2015hik}
want in an improved QE (or 2p2h) model.  It may be interesting to have it coded up as an
alternate, a tool for exploring sensitivities 
(e.g. its a functional form that can be applied to resonance events),
but its definitely not the Valencia QE model.

\begin{figure}[htbp]
\begin{center}
\includegraphics[width=7.0cm]{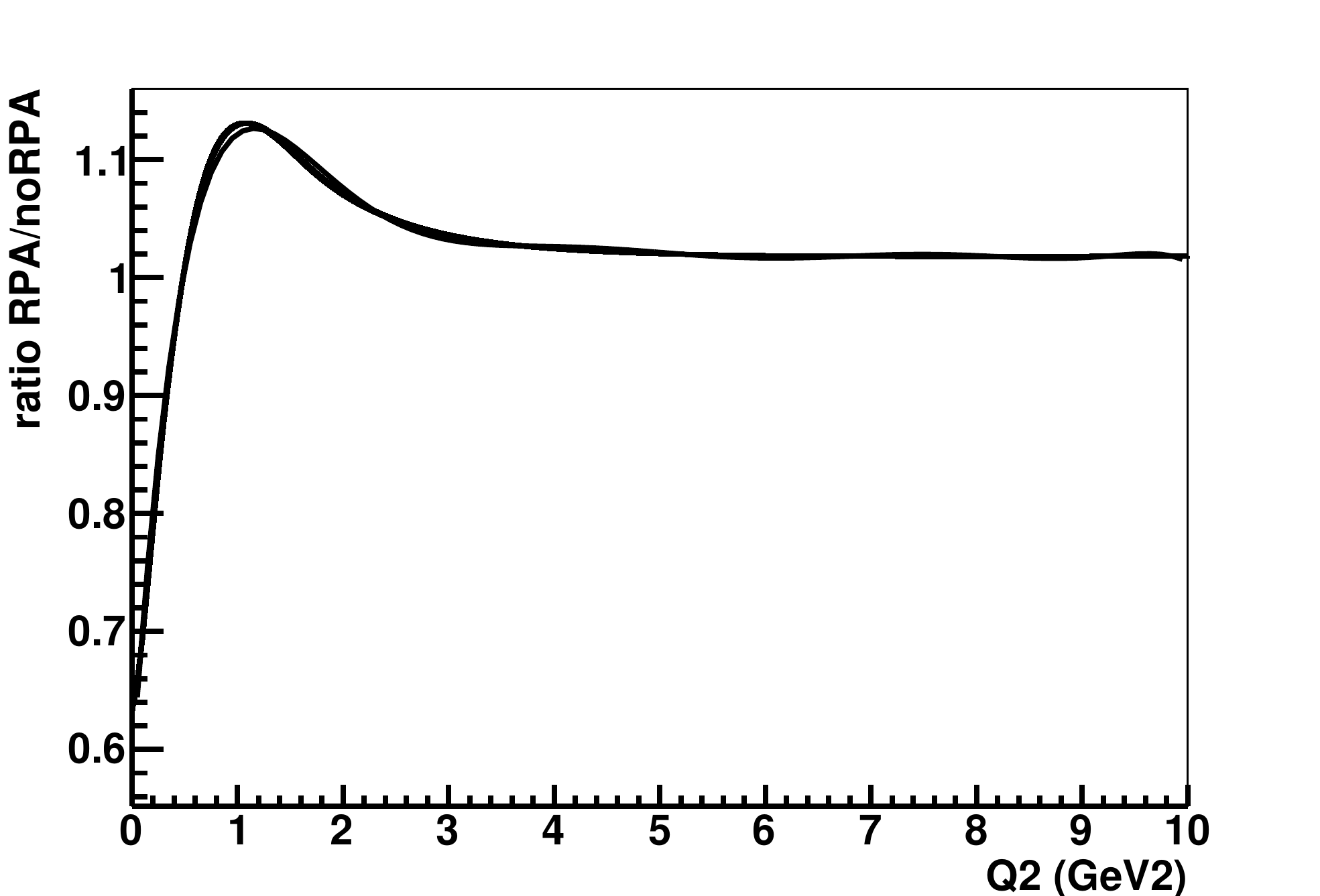}
\includegraphics[width=7.0cm]{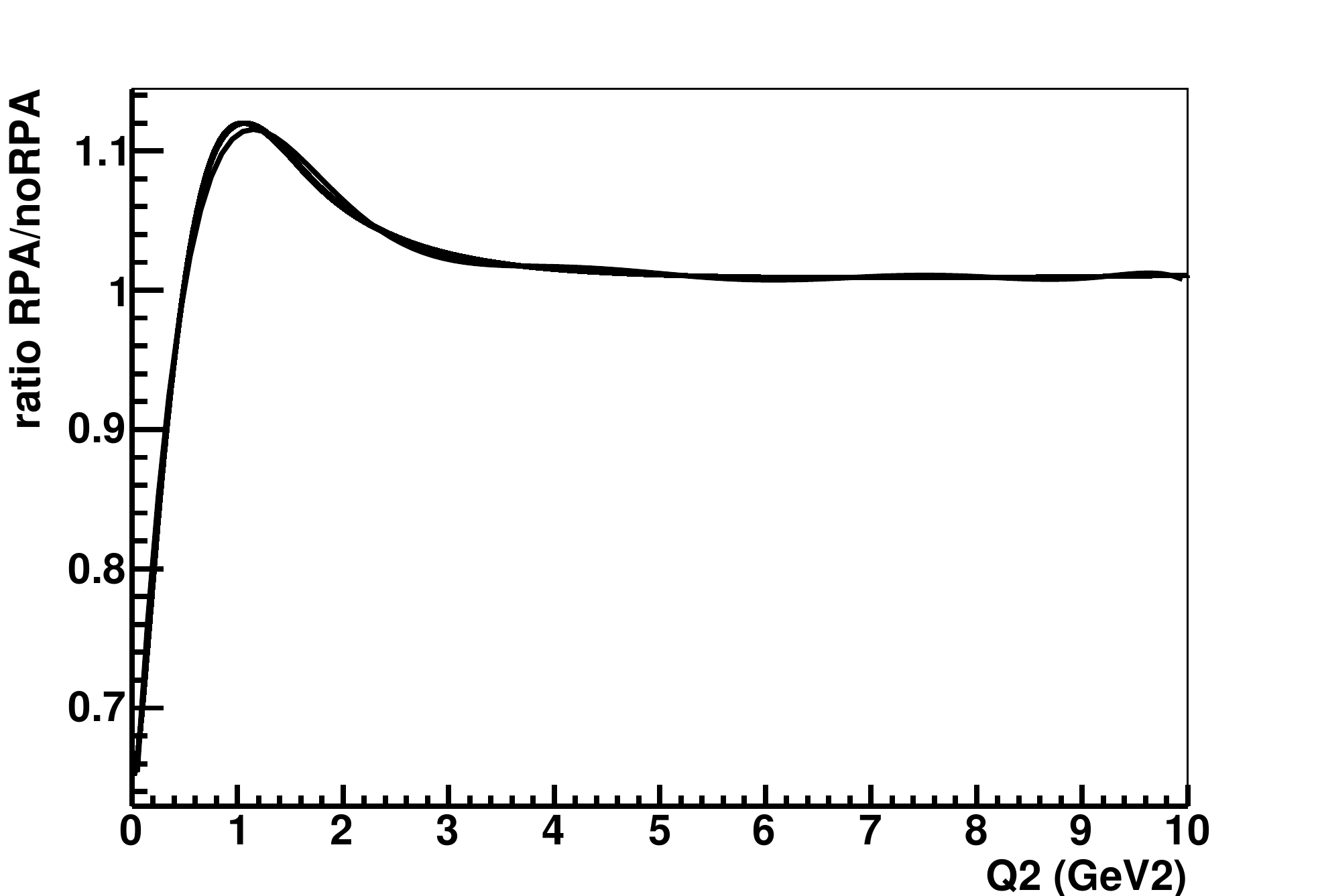}
\parbox{6in}{\caption{Parametrized $r(Q^2)$ directly from the
 Valencia 20 GeV carbon calculation.  Here, a MC method is used to 
integrate the $Q^2$ distribution from the $(q_0,q_3$) calculation,
rather than direct integration from the model.
\label{fig:valenciaq2ratio}}}
\end{center}
\end{figure}

The $r(q^2)$ weights can be obtained directly from the integrated
model, as saved and shown in Fig.~\ref{fig:valenciaq2ratio}, same as
used for the central (red) value in Fig.~\ref{fig:uncertainty}.  In
addition, a polynomial with 10 parameters can be fit, for
computational situations
where reading in a large data file is not efficient enough.  This fit
is also shown on Fig.~\ref{fig:valenciaq2ratio}, where you can see
variations and wiggles that are not larger than 0.5\%, except for the
data points closest to $Q^2 = 0$.  As of this
writing, we package the polynomial fits, the model Q2 ratios, and the
2D ratios in a single file for the neutrino, carbon combination, and a
second file for anti-neutrino, carbon.  Similar files are available
for $^{16}$O, $^{40}$Ar, $^{40}$Ca, $^{56}$Fe, $^{208}$Pb, thus
covering a wide range of nucleus sizes.

\clearpage
\pagebreak

\begin{figure}[htbp]
\begin{center}
\includegraphics[width=5.0cm]{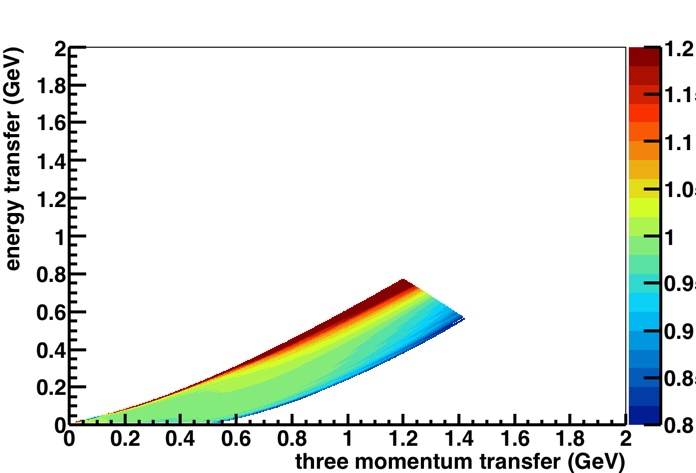}
\includegraphics[width=5.0cm]{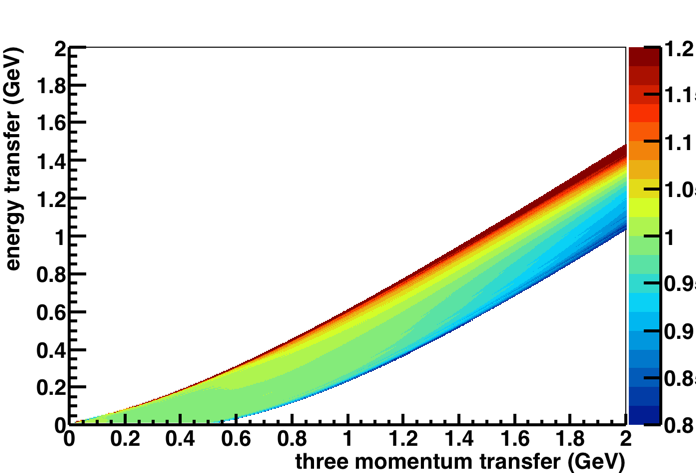}
\includegraphics[width=5.0cm]{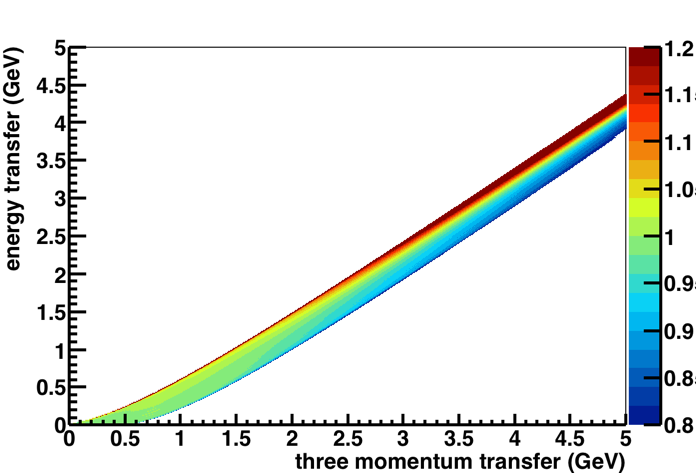}
\parbox{6in}{\caption{Energy dependence illustrated with the
    double ratio (oxygen 1GeV rpa/norpa)/(oxygen
    20GeV rpa/norpa) for neutrino reactions, and the same with 2 GeV
    and 5 GeV in the numerator.  Note, the color scale
    here goes from 0.80 to 1.20.  
    The suppression at very low energy transfers is
    the same, but there is distortion near the upper kinematic
    boundary which moves with it as the energy is increased.
\label{fig:energydependence}}}
\end{center}
\end{figure}

\section{Energy dependence}
\label{sec:energydependence}

The energy dependence of the predicted RPA effect is well controlled for most
analyses at low momentum transfer, for the MINERvA flux.   

The simplest implementation strategy is to use the weights
generated from 20 GeV neutrino interactions for all energies.
The lower energy {\small GENIE} events to be weighted simply
have no events at impossible kinematics.
Events at higher kinematics than available at 20 GeV will get a weight of 1.00$^{+0.03}_{-0.00}$ in
any case.
In the following discussion, I will refer to a the ``energy
simplicity mistake'' to be the difference between result 
from the 20 GeV histogram and 
the one we should have applied if we generated the
weight from the exact energy.
This energy simplicity mistake is within  5\% at most
kinematics, and better than 1\% when integrating slices of $q_3$.

The actual detail of the Valencia model is more interesting, and
is shown using a pair of double ratios in
Fig.~\ref{fig:energydependence}.
These are computed for oxygen just for variety; the results are the
same for carbon.  The left plot is (neutrino oxygen 1 GeV RPA / noRPA) /
(neutrino oxygen 20 GeV RPA / noRPA).  The middle plot is the same, but
with 2 GeV in the numerator, and the right plot is 5 GeV, where the
bounds are wider to show the entire range.  The color axis range is
the same for all three plots, but smaller,
just $\pm$20\% here, instead of $\pm$60\% elsewhere in this paper.

At low momentum transfer, the double ratio is all green.  Applying a
weight based on the 20 GeV calculation accurate to within 5\%, in the
suppression region especially,
smaller than the errors we were already going to assign in that
region.

In contrast, in the upper kinematic regions where the RPA effect
enhances the QE cross section, there is a significant distortion of
the RPA weight, and therefore the energy-transfer spectrum.
At specific kinematics, the energy simplicity mistake (applying the 20 GeV weight) could be as much as
20\% different than we would have applied had our implementation had
built-in access to the full energy dependence.   

\begin{figure}[tbp]
\begin{center}
\includegraphics[width=5.0cm]{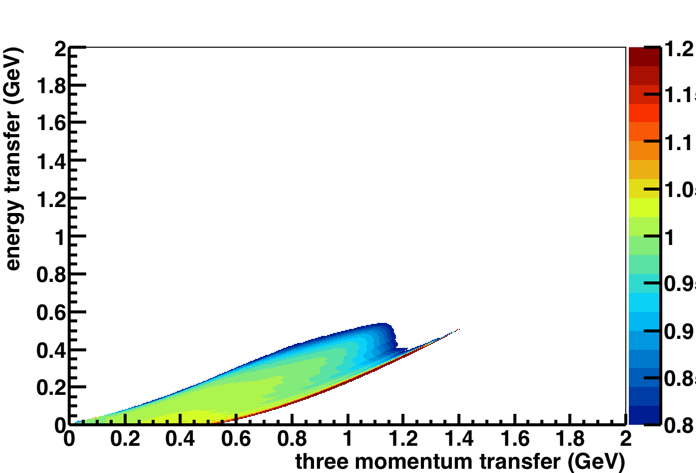}
\includegraphics[width=5.0cm]{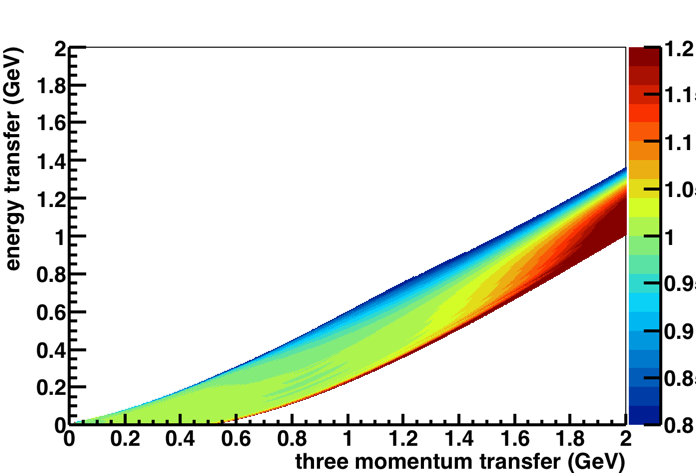}
\includegraphics[width=5.0cm]{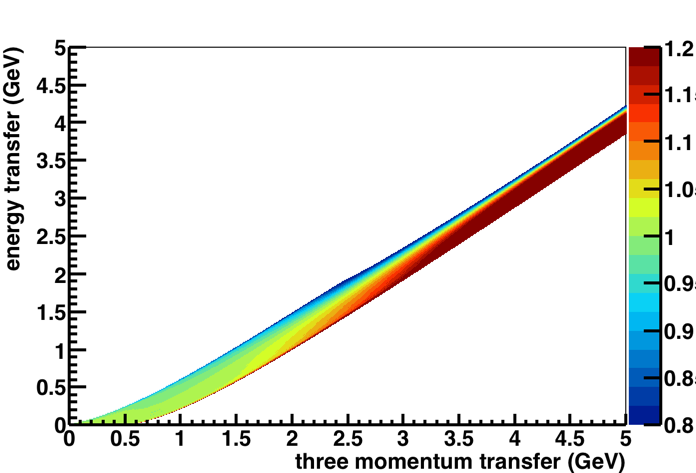}
\parbox{6in}{\caption{Energy dependence illustrated with the
    double ratio (oxygen 1GeV rpa/norpa)/(oxygen
    20GeV rpa/norpa) for anti-neutrino reactions, and the same with 2 GeV
    and 5 GeV in the numerator.  Note, the color scale
    here goes from 0.80 to 1.20.  
    The suppression at very low energy transfers is
    the same, but there is distortion near the upper kinematic
    boundary which moves with it as the energy is increased.
    The behavior at 1 GeV especially (and 2 and 5 GeV somewhat 
involves a more severe kinematic
cutoff (weights are absent from the plot, try flipbook with previous page) along the high energy
transfer side of the distribution, not seen in the neutrino case.
\label{fig:energydependenceanu}}}
\end{center}
\end{figure}

The most serious effect is following the upper kinematic
boundary of the 1 GeV and 2 GeV samples in this case.  It is more
severe, not at fixed kinematics, but for kinematics closest to the
backscattered boundary.   By a typical MINERvA energy like 5 GeV, the energy
simplicity distortion
only becomes significant above $q_3 = 2$ GeV.
That the severity is worse still for the anti-neutrino
case suggests a common cause within the calculation.  The
vector-axial interference term gains relative significance at momentum transfers closer to
the neutrino energy (lepton is more backscattered), 
and especially so for anti-neutrino where the
interference term reduces the cross section.

Though these 20\% effects are visible in the model at specific
high momentum-transfer kinematics, 
they represent a shift in the cross section strength to
lower energy transfers, rather than an overall modification of the
enhancement.
Integrating the cross section along a line of constant three-momentum
transfer yields a net energy simplicity mistake that translates into a
shift in average energy transfer of 5 to 10 MeV, but
at most 1\% of the total event rate.   

We already chose a simplification in Sec.~\ref{sec:modelskew} that takes advantage of this
integral property of the Valencia model.  We transition from the 2D
$r(q_0,q_3)$ weight below $Q^2 = 3.0$ GeV$^2$ to a 1D $Q^2$ based weight,
then to no weight $1.00 \pm 0.05$ above 9.0 GeV$^2$.   The justification
here is the same:  the accumulated simplifications on a small
correction for the higher $Q^2$ QE tail, which anyway we may never be
able to analyze in detail, are safer than the application of
the high momentum transfer 2D weights from the Valencia model to GENIE events.  

This issue must be severe for T2K, MiniBooNE, and MicroBooNE, where all
details of the RPA effect described here are active for their whole QE kinematics all
the time.  MiniBooNE explicitly measures backscattered leptons in all
their core measurements.   For NOvA to use this procedure, it may be enough to choose a
baseline calculation of 3 or 4 GeV.   The relevant energies in DUNE
span the whole range, but they have the luxury of some time to
investigate their best strategy, or even wait for more information
from currently running experiments. 
If MINERvA analysis progresses to
the point where we are picking out
backscattered QE events at $Q^2 = 5$ GeV$^2$
and analyzing detailed hadron energy distributions, 
we will have achieved so many
other victories along the way, we'll be too busy with 
our adoring fans and ``prizes and like that'' \footnote{phrase from a
  1980's Minneapolis radio D.J. that you would be challenged to find with a
name-brand search engine} to remember to fix this up.

\section{Other nuclei}
\label{sec:othernuclei}

This figure illustrates the ratio of ratios Pb(RPA/not)/C(RPA/not).
The Valencia model changes the Q-value for the reactions, as well as
the nuclear density parameters used for the RPA and local Fermi-gas
parts of the model.  

\begin{figure}[htbp]
\begin{center}
\includegraphics[width=9.0cm]{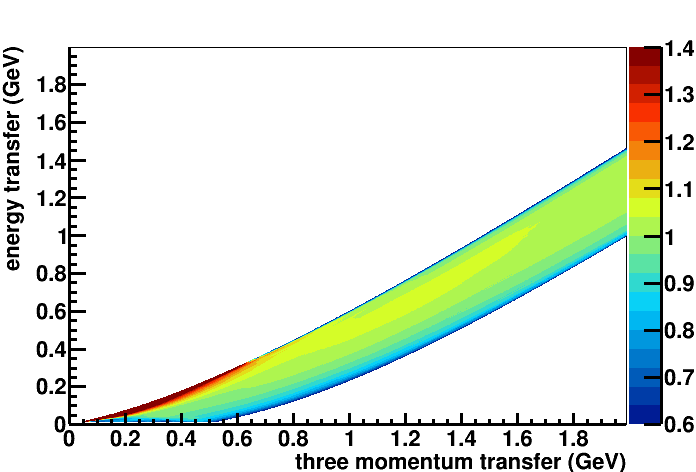}
\parbox{6in}{\caption{The double ratio Pb (RPA/not) / C (RPA/not). 
\label{fig:pbratio}}}
\end{center}
\end{figure}

This also looks like a mild shift in the QE peak to higher energy
transfers for the Pb at very low momentum transfer, and modestly
different shape overall.   
The Q-value piece of the valencia model
allows energy transfers down to near zero for Pb, instead of the 14 or
17 for Carbon.
Combined with an RPA distortion that might otherwise be different
in a larger nucleus.


\clearpage
\pagebreak

\begin{figure}[htbp]
\begin{center}
\includegraphics[width=7.0cm]{figures/nugenieratio2D.png}
\includegraphics[width=7.0cm]{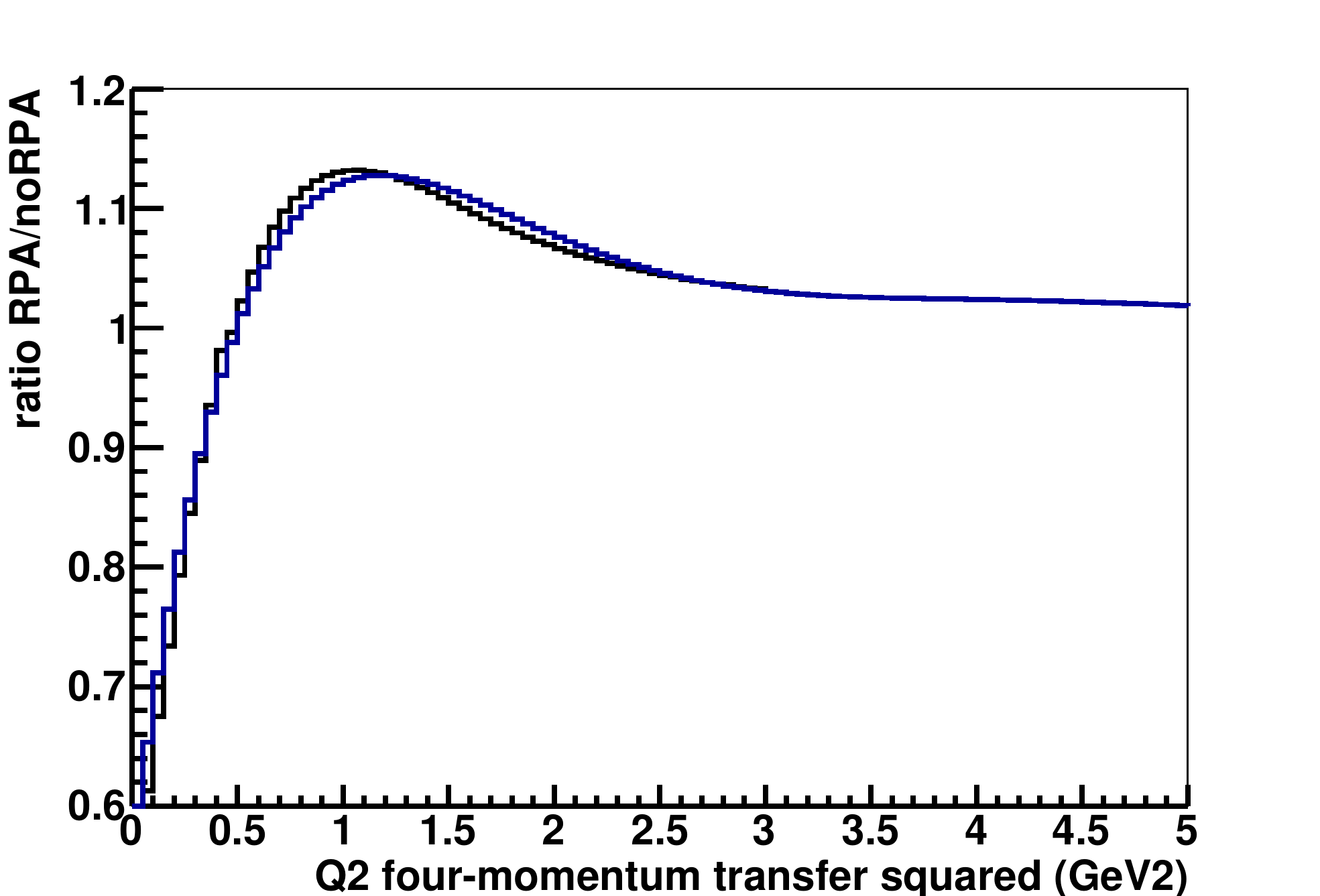}
\parbox{6in}{\caption{Ratio of RPA/noRPA for {\small GENIE} neutrino events with the
    full central-weighting described in the paper.  The 2D weight as
    directly produced by the coded procedure (left) and the resulting ratio after
    the {\small GENIE} events are binned in true $Q^2$ for the complete
    procedure (black) and for the r($Q^2$) parameterization only
    (blue).
   The differences between black and blue are 
    smaller than the uncertainties we already plan to assign, and are
    due to GENIE having a different
    baseline QE model than the Valencia.
\label{fig:closurenuratio}}}
\end{center}
\end{figure}

\section{Closure tests}

At the top of {\em this page}, Fig.~\ref{fig:closurenuratio} 
on the left are the final 2D weights applied to {\small GENIE} neutrino events.
On the right are the same QE {\small GENIE} events integrated to produce the
$Q^2$ distribution, then the RPAweighted/default {\small GENIE} ratio
is shown.  This page can be flipbooked with the antineutrino
equivalent on the next page.

The central-value $Q^2$ ratio comes out nearly identical, so the procedure 
robustly implements the RPA effect as a weight, despite the differences
between the Valencia and {\small GENIE} QE models.  The differences at low $Q^2$ are
overwhelmed by the enhanced uncertainty from muon capture.
The glitch at $Q^2 > 3.0$ GeV$^2$ is due to the model skew applying
Valencia weights to the {\small GENIE} base model events, and is only
1\%.
That is easily within the 3\% uncertainty assigned there,
which anyway will be in quadrature with the larger axial form factor uncertainty.

\clearpage
\pagebreak

\begin{figure}[htbp]
\begin{center}
\includegraphics[width=7.0cm]{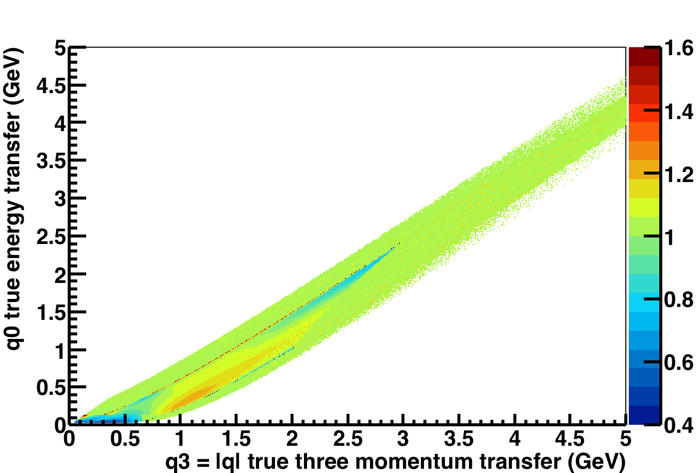}
\includegraphics[width=7.0cm]{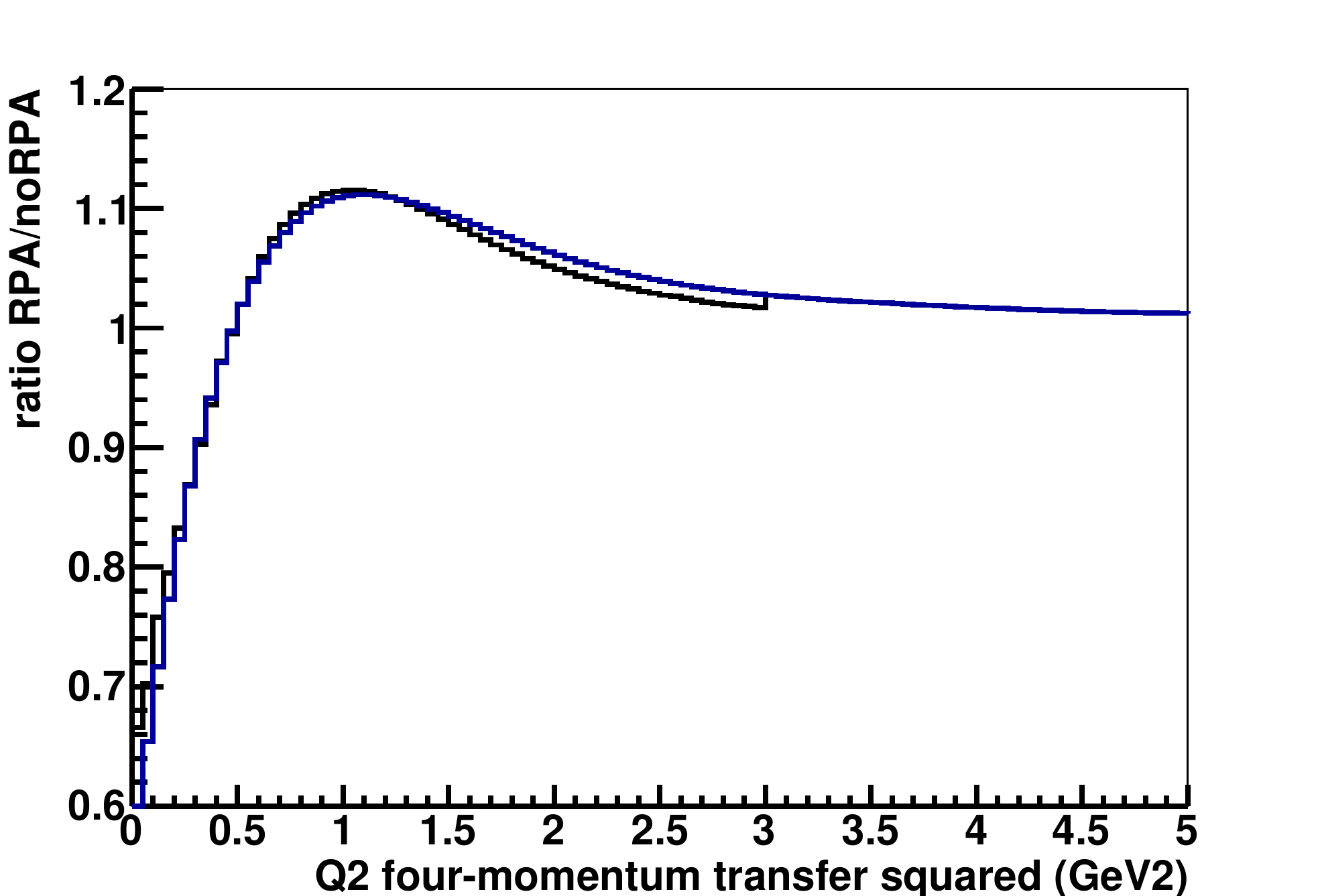}
\parbox{6in}{\caption{Ratio of RPA/noRPA for {\small GENIE} neutrino events with the
    full central-weighting described in the paper.  The 2D weight as
    directly produced by the coded procedure (left) and the resulting ratio after
    the {\small GENIE} events are binned in true $Q^2$ for the complete
    procedure (black) and for the r($Q^2$) parameterization only
    (blue).  The differences between black and blue are 
    smaller than the uncertainties we already plan to assign, and are
    due to GENIE having a different
    baseline QE model than the Valencia.
   In the anti-neutrino case, the base model peaks are more offset, so
   it leads to a small
   step-like artifact.
\label{fig:closureanuratio}}}
\end{center}
\end{figure}

\subsection*{Closure tests anti-neutrino}

At the top of this page, Fig.~\ref{fig:closureanuratio} 
on the left are the final 2D weights applied to {\small GENIE} anti-neutrino events.
Can flipbook in the electronic version of this document.

This also passes the test, that the 2D weight returns the 1D $Q^2$
suppression more accurately than the rest of the uncertainties we will
assign, despite the model skew between {\small GENIE} and Valencia.


\clearpage
\pagebreak

\section{Additional comments}

\subsection{Apply to the resonances?}

Why not ?  
We can, almost. 
But the ($q_0$,$q_3$) weight is inappropriate as is.
The Delta is offset to higher energy and momentum transfers, for the
same $Q^2$, compared to QE events.
We could simply apply the $Q^2$ weights and see what happens. 
I've done that with some generator studies.  The result is a 
fair match for the published MINOS result \cite{Adamson:2014pgc}, 
the MiniBooNE result
\cite{AguilarArevalo:2010bm,AguilarArevalo:2010xt}, 
(and seems to go the right direction to describe some MINERvA distributions too).
A central value like this with an on/off as uncertainty may be
appropriate.

Speculation: a real RPA calculation may reveal a ($q_0$,$q_3$) structure
similar to the QE case, with simple offsets to map it onto a line of
invariant mass W=1.232 GeV, at least for the $\Delta$1232 resonance.  
This would preserve a suppression on the
low $q_0$ side of the W line, and have less suppression on the high
side.  Equally valid guess is we learn from our theory colleagues to 
expect much less RPA suppression for resonance events, but a yet
different effect is large.

\subsection{Neutral current}
The neutral current case is not currently
considered, but should also show the effect, as was there also for the
electron scattering case.  It may be possible to use the simpler $Q^2$ or
the more expensive ($q_0$,$q_3$) version
of this weight, in the code fragment above, or a variation of it, as a
crude approximation for NC events.  Suitable to test for analysis
sensitivity anyway, e.g. for sterile neutrino oscillations analyses.

\subsection{Local Fermi gas}

The Valencia calculation includes a local Fermi gas model (LFG), which is an
additional interesting feature of the model.   You can see the effects
in the 1D plots of {\small GENIE} vs. Valencia, such as
Fig.~\ref{fig:oneDplots}, for which the 500 MeV version is shown
here.   

\begin{figure}[htbp]
\begin{center}
\includegraphics[width=8.0cm]{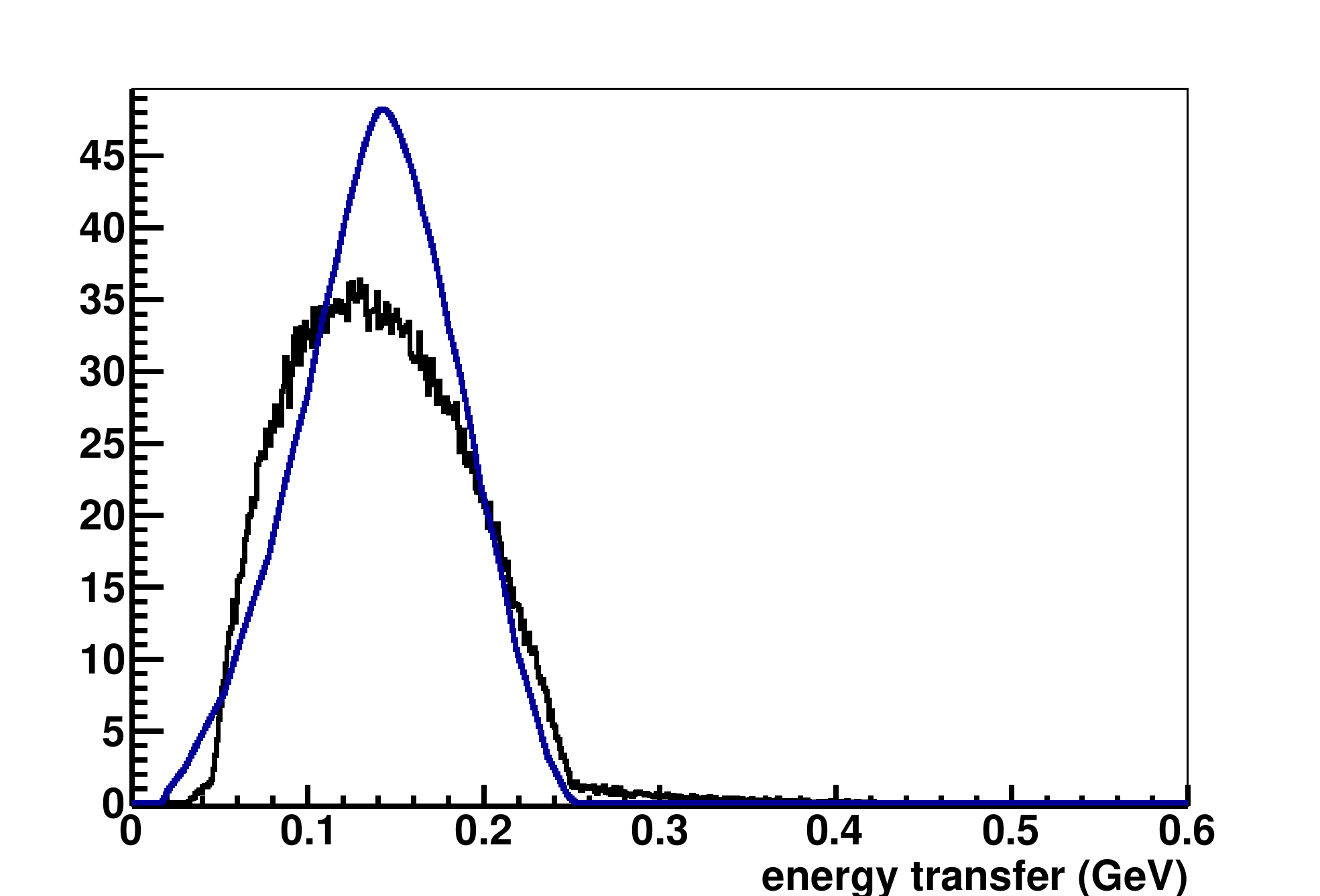}
\includegraphics[width=6.0cm]{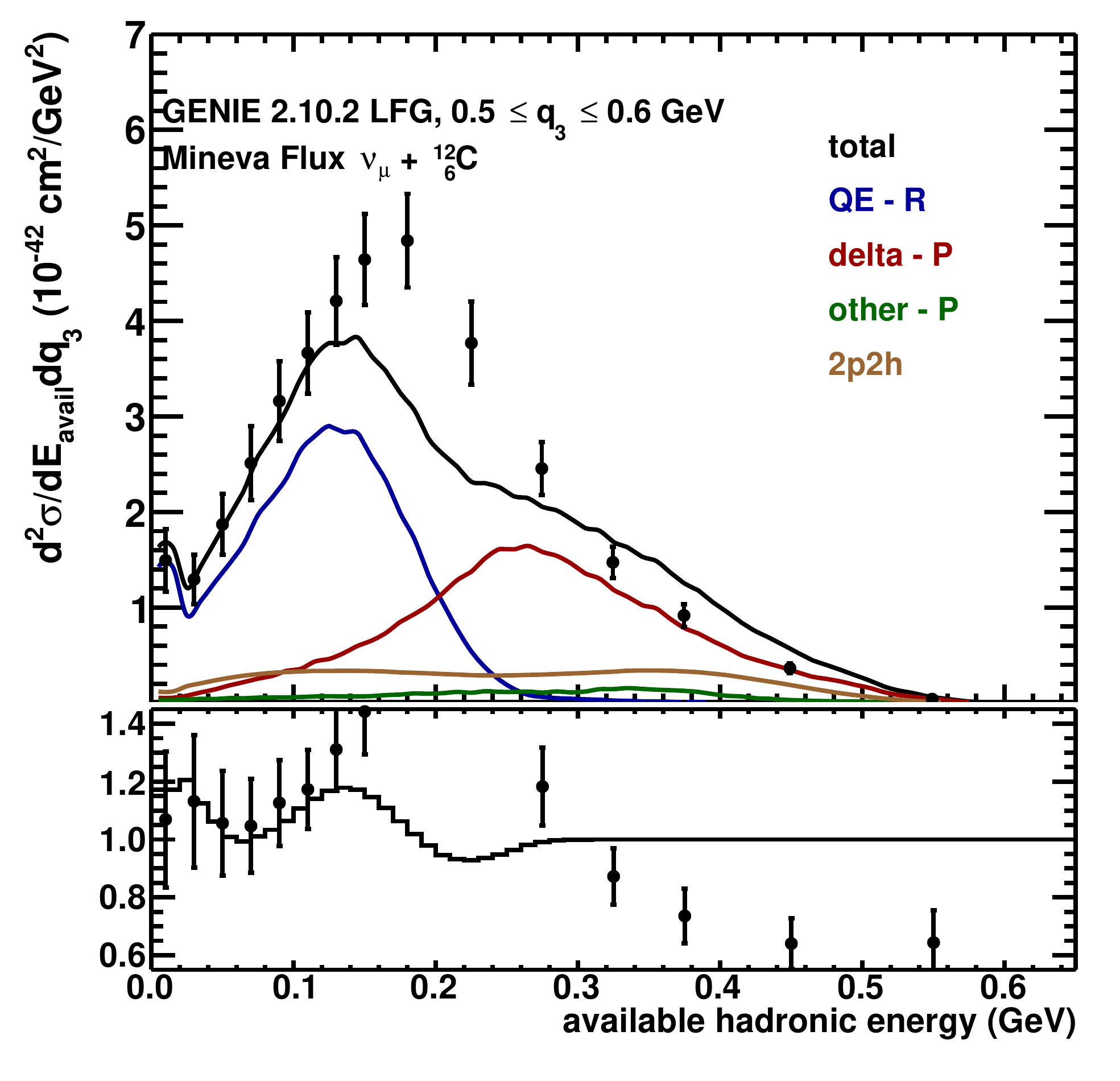}
\parbox{6in}{\caption{Left plot is a variation of earlier plots showing the distortion of
    the cross section in a slice of $q_3=0.5$ GeV due to the local
    Fermi gas.  Both the {\small GENIE} (black) and Valencia (blue)
    distributions have RPA applied.   The integral of the two
    distributions are only different by 6\%, despite such different shapes.
On the right is what happens when the combined weight of the local Fermi
gas is added to the RPA weight, and all the other processes in {\small GENIE}, and compared to data from 
\cite{Rodrigues:2015hik}.  The ratio shows only the change when adding
the local Fermi gas.  [Right side plot by Jake Leistico.]
\label{fig:localfermigas}}}
\end{center}
\end{figure}

We have separately been
exploring a reweight of Valencia LFG to {\small GENIE} QE, applied using the
same histogram 2D weight method.  To leading order,
it appears the two effects are separable.  One example is the right
hand plot, comparing a modified model to the data from
\cite{Rodrigues:2015hik}.   The baseline in that plot already had a
prototype version of the  RPA effect weight described in this paper,
to which a new LFG weight is added based on the 5 GeV Valencia calculation.
It is being compared to the cross section that additionally has a
2D histogram-based weight for the Valencia local Fermi gas.
The distortions are the same size as the uncertainties on the data;
modern experiments and analysis techniques do have sensitivity to these nuclear effects.
However, the local Fermi gas does not distort the model enough to meet
the data, it must be only one component of a post-Fermi gas answer.

Continuing this exploration is
beyond the scope of this paper,
but the reweight technique used for the RPA and here for LFG
may be advantageous for fast exploration of these effects.
An implementation of the full Valencia model, including the local
Fermi gas, is newly available in
{\small GENIE} version 2.12 .

\section{Conclusion}

We have captured the leading order effects of an RPA-type
multi-nucleon model and assigned an uncertainty to these effects.
This procedure can be used with a Fermi gas model from an event
generator to apply an RPA model correction and its uncertainty
as a reweight to already generated events.

\section*{Acknowledgements}

We are grateful for the work of Federico Sanchez and his student
Bruno [] .  Juan Nieves offered suggestions to improve the paper, and
insight from the latest preprint with Joanna Sobczyk.
The Valencia model, with RPA and Local Fermi Gas effects has
also now been coded directly into {\small GENIE} by Steve Dytman and his
student , and is available as of release 2.12.  The parameters
discussed here may be exposed in that implementation, I'm not sure.
That version was not
yet available for testing for this analysis, but will be of interest
to some users of CCQE interaction models.  
This work was supported by NSF awards 1306944 and 1607381 to the
University of Minnesota Duluth.

\bibliography{rpasyst}

\end{document}